\documentclass[aip,jcp,twocolumn,reprint,superscriptaddress,longbibliography]{revtex4-1}
\usepackage[usenames, dvipsnames]{color}
\usepackage{graphicx}
\usepackage{color}
\usepackage{epstopdf}
\usepackage{epsfig}
\usepackage{bm}
\usepackage{xcolor}
\usepackage{amsmath}
\usepackage{amsfonts}
\usepackage{float}
\usepackage{subfigure}
\usepackage{verbatim}
\usepackage[version=3]{mhchem}
\usepackage[utf8]{inputenc}
\usepackage{epstopdf}
\usepackage{scalerel}
\usepackage{multirow}
\usepackage{booktabs}
\usepackage{hyperref}
\usepackage[para,flushleft]{threeparttable}

\setcitestyle{super}

\newcommand{\rb}{\mathbf{r}}

\newcommand{\avg}[1]{\left<#1\right>}
\newcommand{\len}[1]{\left|#1\right|}
\newcommand{\brac}[1]{\left[#1\right]}
\newcommand{\curly}[1]{\left\{#1\right\}}
\newcommand{\para}[1]{\left(#1\right)}

\newcommand{\kT}{\ensuremath{k_{\rm B}T}}
\newcommand{\Mb}{\ensuremath{\mathcal{M}}}

\begin{document}


\title{Electronic Fluctuations and Ionic Dynamics in Molten Silver Iodide}



\author{Harender S. Dhattarwal}
\author{Richard C. Remsing}
\email[]{rick.remsing@rutgers.edu}
\affiliation{Department of Chemistry and Chemical Biology, Rutgers University, Piscataway, NJ, 08854, USA}




\begin{abstract}
Molten salts are high-temperature ionic liquids whose unique combination of strong Coulombic interactions, large polarizabilities, and high ionic conductivities makes them central to energy storage, metallurgy, and nuclear technology.
Understanding their delicate balance of Coulomb forces, short-range repulsion, and electronic polarization, particularly regarding the role that electronic fluctuations play in their structure and dynamics, is critical to predictively designing molten salts for applications of interest. 
We investigate the importance of electronic fluctuations in molten AgI using density functional theory, a universal machine learning model (Orb), and a classical, empirical pairwise model of interionic interactions. 
We find that directional polarization fluctuations of iodide ions enhance Ag$^+$ diffusion, manifesting as enhanced force fluctuations and structure in the time-dependent friction experienced by the cations. 
The coupling between iodide polarization fluctuations and silver diffusion creates a dynamic asymmetry; Ag$^+$ motion is tightly linked to the instantaneous polarization of neighboring I$^-$, whereas I$^-$ dynamics are relatively unperturbed by electronic fluctuations.
For all structural and dynamic quantities investigated, the Orb model is in excellent agreement with density functional theory-based simulations, highlighting the ability of this universal neural network potential to capture many-body polarization effects. 
In contrast, the empirical force field fails to reproduce key structural and dynamic quantities involving cations, ultimately because it neglects dynamic electronic fluctuations. 
Our findings connect liquid-state ionic dynamics with the ``electronic paddle-wheel'' mechanism of ionic diffusion in superionic solids and motivate further exploration of polarization fluctuation effects in complex electrolytes and ionic liquids.
\end{abstract}


\maketitle

\raggedbottom

\section{Introduction}
Molten salts and high-temperature ionic liquids often contain highly polarizable ions, and induced electronic polarization can dramatically influence their structure and dynamics~\cite{Salanne2011polarization,Madden1993MgCl2,Porter2022,borodin2009polarizable,Bedrov2019,bedrov2010influence}. 
Induced dipole and quadrupole polarization are essential to capturing key properties of ionic systems,
such that adding anion polarization to rigid-ion models more accurately predicts structural features that are missed by simpler models~\cite{Madden1996quadrupole}.
Therefore, many-body electronic effects must be included to accurately model ionic fluids~\cite{madden1996covalent}.
Molten silver iodide serves as an ideal model molten salt to study polarization effects, because it is composed of a hard cation and a highly polarizable anion.
Models of AgI with a polarizable iodide reproduce the first-sharp diffraction peak ($\sim1$~\AA$^{-1}$) observed in neutron scattering, and strongly preserve its ``superionic'' character; Ag$^+$ ions diffuse much faster than I$^-$ ions~\cite{Trullas2008polarization}. 
These polarizable models also predict ionic conductivities closer to experimental values than rigid-ion models.
As a result, anionic polarization is crucial to capture the asymmetric dynamics and transport of molten AgI.
AgI is also a canonical solid-state superionic conductor and exhibits analogous polarization-driven mechanisms. 
Cation diffusion in $\alpha$-AgI involves an ``electronic paddle-wheel'' mechanism, in which rotations of I$^-$ electron density couple to and facilitate Ag$^+$ diffusion~\cite{Dhattarwal2024paddle}. 
The strong coupling between local electronic polarization and cation motion can provide a unified picture of ion transport across a wide range of ionic materials.
Here we bridge these liquid- and solid-state perspectives by examining the connection between electronic fluctuations and ionic dynamics in molten AgI at 1600~K. 
We illustrate the impact of electronic fluctuations on ionic dynamics by comparing a model without polarization fluctuations --- the classical rigid-ion force field of Vashishta and Rahman~\cite{AgI_Vashishta} --- to models that include a description of electronic fluctuations --- density function theory (DFT) and the Orb univeral neural network potential~\cite{Orb2025}. 
After investigating the structure of AgI produced by the three models, we explicitly probe how instantaneous electronic fluctuations influence ionic motion by analyzing velocity and force autocorrelation functions within a generalized Langevin equation framework. 
In this approach, we also examine the time-dependent memory or friction kernel, which provides a quantitative measure of how past force fluctuations influence an ion's dynamics.
Our results reveal a pronounced cation-anion asymmetry in molten AgI; Ag$^+$ diffusion is strongly affected by directional polarization fluctuations of I$^-$, whereas I$^-$ dynamics are relatively unaffected. 
The coupling between electronic fluctuations and cation dynamics manifests as enhanced structure in the force autocorrelation and time-dependent friction on Ag$^+$. 
Importantly, the Orb model reproduces the DFT results for all structural and dynamic properties, underscoring its accuracy in capturing many-body polarization effects. 
In contrast, the classical force field (FF) model fails to reproduce key features in both structure and dynamics involving cations, highlighting the importance of explicit electronic interactions.
By quantifying memory functions and force fluctuations, our work highlights their ability to probe electronic polarization effects in liquid ionic conductors. 
The observed liquid-state polarization dynamics closely mirror the electronic paddle-wheel mechanism in solid electrolytes, and suggest that analogous effects may occur in other ionic liquids and electrolytes.
These findings motivate further exploration of fluctuating polarizability in complex ionic materials.

\section{Theory and Simulation Details}

We analyzed the structure and dynamics of molten AgI at 1600~K from density functional theory-based molecular dynamics (DFT-MD) simulations and compared it with molecular dynamics simulations performed using a universal interatomic potential, Orb~\cite{Orb2025}, and the Vashishta-Rahman empirical pair potential~\cite{AgI_Vashishta}. 
DFT-MD simulations were performed using the CP2K software package~\cite{CP2K2020} with the PBE exchange-correlation functional~\cite{PBE1996} and Goedecker-Teter-Hutter (GTH) pseudopotentials~\cite{GTH1996,Krack2005}. We employed the molecularly optimized double-$\zeta$ with valence polarization (DZVP) basis set with cutoffs of 450~Ry for plane wave energy and 60~Ry for the reference grid. 
The machine learning potential-based MD simulations were performed using the pre-trained Orb-v3-conservative-inf model~\cite{Orb2025}. 
Since the structure and dynamics of molten AgI was well captured by the pre-trained Orb model, we did not perform any additional fine tuning. 
We used the Vashishta-Rahman potential for the classical force field-based MD simulations~\cite{AgI_Vashishta,dhattarwal2025electronic}.
The full form of the potential and relevant parameters can be found elsewhere~\cite{AgI_Vashishta,dhattarwal2025electronic}, and this pairwise interatomic interactions include contributions from repulsive cores, dispersion, Coulomb, and polarization interactions between ions. 
We note that the charges on Ag$^+$ and I$^-$ are fixed at $0.6e$ and $-0.6e$, respectively, 
and only the iodide ions have a non-zero electronic polarizability.
The polarization term (which scales as $r^{-4}$) is non-zero for only I-I and Ag-I interactions and neglects Ag$^+$ polarization and any feedback between Ag$^+$ and I$^-$ induced polarization,
such that the polarization term accounts for the averaged pairwise effects of iodide polarization. 
We ran DFT-MD simulations with a cubic simulation cell containing 108~atoms and a box length of 15.2565~\AA with an initial configuration in the $\alpha$-AgI structure.
This was heated to 1600~K and equilibrated for 10~ps in the canonical ensemble (NVT) using the canonical velocity rescaling (CSVR) thermostat~\cite{CSVR2007}. 
The system was further equilibrated for 80~ps in microcanonical ensemble (NVE) with a time step of 1~fs. 
The last 60~ps of the NVE trajectory was used for analysis. 
In order to capture long range structural ordering, we performed another set of MD simulations of a larger system containing 864 atoms (432 Ag$^+$ and 432 I$^-$) in a cubic box of length 30.513~\AA.
The Orb MD simulations were performed using the Atomic Simulation Environment (ASE)~\cite{ASE}. Initial NVT equilibration was performed using Langevin dynamics for 2~ns with a timestep of 1~fs. 
The friction coefficient was set to 0.1~fs$^{-1}$. 
The FF MD simulations were performed using the LAMMPS software package~\cite{LAMMPS}. 
For the FF model, the initial NVT equilibration was performed using Nos\'{e}-Hoover thermostat for 2~ns with a timestep of 1~fs~\cite{Nose1984,Hoover1985}. 
For both of these models, the system was further equilibrated for 200~ps in NVE ensemble with a timestep of 0.5~fs and the last 100~ps of the NVE trajectory was used for analysis.

\section{Results}

\subsection{Structure}

\begin{figure}[tb]
\begin{center}
\includegraphics[width=0.49\textwidth]{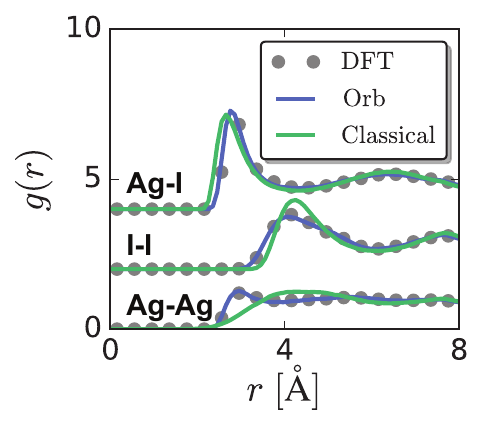}
\end{center}
\caption
{Pair correlation functions, $g(r)$, for Ag$^+$-Ag$^+$, I$^-$-I$^-$, and Ag$^+$-I$^-$ pairs computed from DFT, Orb, and FF models.}
\label{fig:rdfs}
\end{figure}

\begin{figure*}[tb]
\begin{center}
\includegraphics[width=0.85\textwidth]{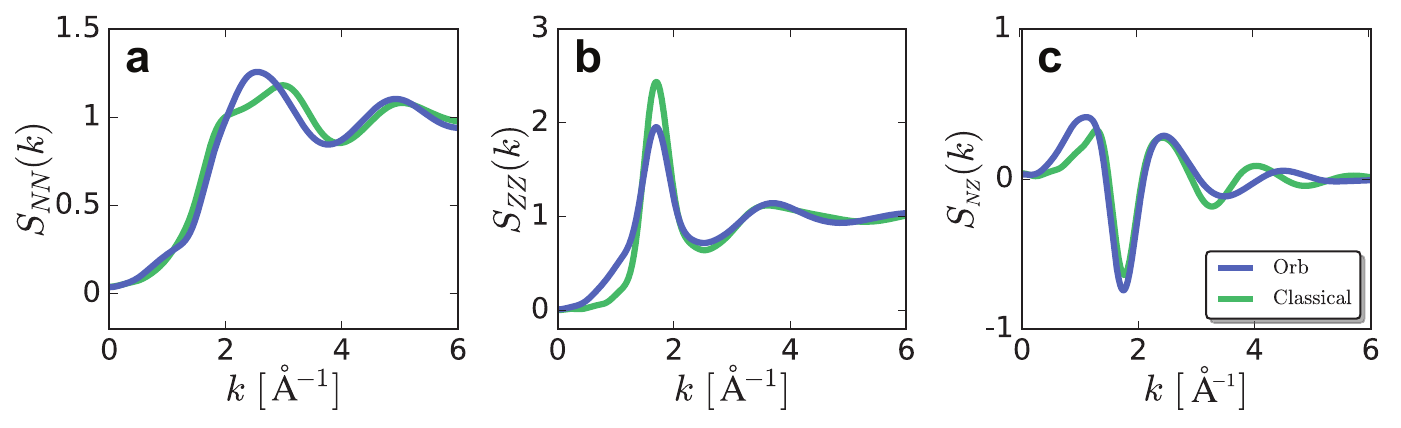}
\end{center}
\caption
{(a) Density-density structure factor, $S_{NN}(k)$, and (b) charge-charge structure factor, $S_{ZZ}(k)$, for molten AgI obtained from Orb and FF models.
}
\label{fig:sk}
\end{figure*}

\begin{figure*}[tb]
\begin{center}
\includegraphics[width=0.85\textwidth]{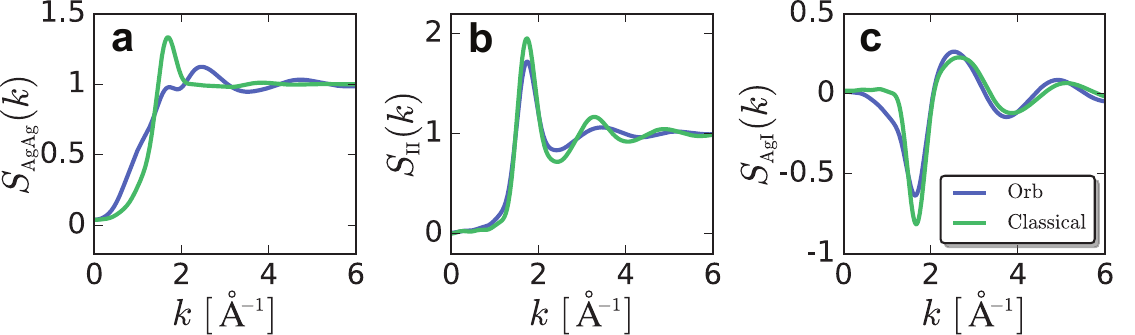}
\end{center}
\caption
{Partial structure factors for (a) Ag-Ag, (b) I-I, and (c) Ag-I correlations in molten AgI as predicted by the Orb and FF models.
}
\label{fig:psk}
\end{figure*}

%
The structure of molten AgI at 1600 K was characterized through site-site pair distribution functions, $g(r)$, for Ag$^+$-Ag$^+$, I$^-$-I$^-$, and Ag$^+$-I$^-$ pairs, Fig.~\ref{fig:rdfs}. 
The Orb model produces site-site pair distribution functions in nearly exact agreement with the ab initio DFT results for all site-site correlations. 
The FF model captures the general features of the anion-anion and cation-anion site-site distribution functions,
however, it produces a cation-cation $g(r)$ with significantly less structure than the Orb and DFT models. 
The cation-cation $g(r)$ from Orb and DFT indicate that the Ag$^+$ ions are strongly correlated, evidenced by a peak in $g(r)$ at small distances.
However, the FF model does not capture this enhanced correlation, which is a manifestation of polarization-induced screening of cation-cation repulsion~\cite{Trullas2008polarization}, and similar effects have been observed in other molten salts~\cite{chahid1998structure,bitrian2006molecular}.
The FF model includes a description of the average effects of I$^-$ polarization on pairwise interaction through an explicit polarization term, suggesting that the inaccuracies in the FF cation-cation $g(r)$ originate from a lack of dynamic directional electronic fluctuations that lead to many-body interactions. 
Such many-body interactions are included in both Orb and DFT and are crucial to describing the structure and dynamics in the solid-state superionic phase of AgI as well~\cite{dhattarwal2025electronic}. 
The FF model is able to describe the I$^-$-I$^-$, and Ag$^+$-I$^-$ site-site $g(r)$ well because they do not require as accurate a description of electronic fluctuations. 
The Ag$^+$ ions are hard and essentially nonpolarizable, such that complex many-body effects can be neglected to a good approximation.
As a result, the FF Ag$^+$-I$^-$ $g(r)$ is in excellent agreement with DFT and FF model, though the first peak is shifted to slightly smaller distances, most likely the result of inaccurate repulsive interactions between ionic cores. 
The Ag$^+$-I$^-$ $g(r)$ indicates that each iodide is surrounded by a coordination shell of silver cations, such that the I$^-$-I$^-$ interactions are mediated by nonpolarizable ions. 
However, many-body interactions can still be significant in these indirect interactions, such that the I$^-$-I$^-$ $g(r)$ is only in qualitative agreement with DFT and FF model. 
The first peak is too large and too narrow, which also a result of neglecting many-body interactions arising from directional electronic effects. 
We further characterized the structure of molten AgI through the total structure factor, $S_{NN}(k)$, given by
\begin{equation}
S_{NN}(k) = \frac{(b_{{\rm Ag}}^2 S_{{\rm AgAg}}(k) + b_{{\rm I}}^2 S_{{\rm II}}(k) + 2b_{{\rm Ag}}b_{{\rm I}}S_{{\rm AgI}}(k))}
{(b_{\rm Ag}^2 + b_{\rm I}^2)},
\end{equation}
where $b_{\rm Ag}$ and $b_{\rm I}$ are the neutron scattering lengths of Ag and I, respectively, and $S_{\alpha \beta}(k)$ is the partial structure factor between species $\alpha$ and $\beta$, 
\begin{equation}
S_{\alpha \beta}(k) = \delta_{\alpha \beta} + \rho_N \sqrt{x_{\alpha}x_{\beta}} \int \para{g_{\alpha \beta}(r)-1} \frac{{\rm sin} kr}{kr} 4 \pi r^2 dr.
\end{equation}
Here $\delta_{\alpha \beta}$ is the Kronecker delta function, $\rho_N$ is the number density, and $x_{\alpha}$ is the molar fraction of species $\alpha$.
In addition to the density-density structure factor, $S_{\rm NN}(k)$, we also computed the charge-charge structure factor, $S_{ZZ}(k)$, 
\begin{equation}
S_{ZZ}(k) = \frac{1}{2}\brac{q^2 S_{\rm AgAg}(k) + q^2 S_{\rm II}(k) - 2 q^2 S_{\rm AgI}(k)},
\end{equation}
where $q=q_{\rm Ag} = -q_{\rm I}$ is the magnitude of the charge on each ion,
and we used $q=1$ to compare the Orb and FF results with a similar weighting.
Accurately computing structure factors to low wavevectors requires large system sizes, and we therefore only examine the structure factors for the Orb and FF models, working under the assumption that Orb accurately predicts the structure of the AIMD simulations. 
Polarization effects are known to create a prepeak in the density-density and charge-charge structure factors~\cite{Trullas2008polarization}. 
Indeed, $S_{\rm NN}(k)$ and $S_{\rm ZZ}(k)$ exhibit additional features below first peak near $k=2$~\AA$^{-1}$, Fig.~\ref{fig:sk}.
However, the higher temperature studied here (1600~K versus 923~K in previous work~\cite{Trullas2008polarization}) smears out the prepeak into a shoulder in each structure factor.
Such a shoulder is lacking in the structure factors produced by the FF model. 
Cation-cation and anion-anion ordering can be examined through the density-charge structure factor, $S_{NZ}(k)$, given by
\begin{equation}
S_{NZ}(k) = \frac{1}{2}\brac{q S_{\rm AgAg}(k) - q S_{\rm II}(k)}.
\end{equation}
Positive and negative peaks in density-charge structure factor correspond to contributions from cation-cation and anion-anion correlations, respectively. 
For the Orb model, $S_{NZ}(k)$ exhibits a significant positive peak at  $k\approx1.1$~\AA$^{-1}$, suggesting that cation-cation correlations contribute to the prepeak in the structure factors.
The first peak in the $S_{NZ}(k)$ of the FF model is less pronounced and shifted to higher $k$,
reflecting the inability of this model to capture longer-range cation-cation correlations, which is also consistent with the lack of structure in the FF cation-cation $g(r)$.
The difference between the Orb and FF models demonstrate that the cation-cation correlations that give rise to the prepeak are sensitive to polarization fluctuations. 
The origin of the prepeak can be discerned from examination of partial structure factors, Fig.~\ref{fig:psk}. 
Partial structure factors involving cations, $S_{\rm AgAg}(k)$ and $S_{\rm AgI}(k)$, are significantly different when electronic fluctuations are included. 
Both $S_{\rm AgAg}(k)$ and $S_{\rm AgI}(k)$ have a shoulder at low $k$, and $S_{\rm AgAg}(k)$ exhibits two peaks at $k\approx 1.75$~\AA$^{-1}$ \ and $k\approx 2.5$~\AA$^{-1}$, corresponding to two structural length scales, in addition to the shoulder.
In contrast, the FF model $S_{\rm AgAg}(k)$ lacks a shoulder at low $k$ and only has a single peak at $k\approx1.75$~\AA$^{-1}$, consistent with the single broad peak in the cation-cation $g(r)$ produced by the FF model. 
Similarly, the shoulder at small $k$ in $S_{\rm AgI}(k)$ is only observed in the Orb model and not the FF model. 
However, $S_{\rm II}(k)$ is qualitatively similar for both models, with peaks and minima at nearly the same wavevectors. 
These results suggest that polarization effects mainly impact the silver cations and their structuring, and have little impact on the structuring of the iodide ions. 
This can be understood through the ability of electronic fluctuations on iodide to screen interactions between cations~\cite{Trullas2008polarization}, allowing them to approach more closely than in the absence of such fluctuations (in the FF model, for example). 
To the extent that structure dictates dynamics, we may anticipate that cation dynamics will be impacted most by the inclusion of electronic fluctuations, while anion dynamics will exhibit minimal changes, as discussed further below.

\begin{figure}[tb]
\begin{center}
\includegraphics[width=0.45\textwidth]{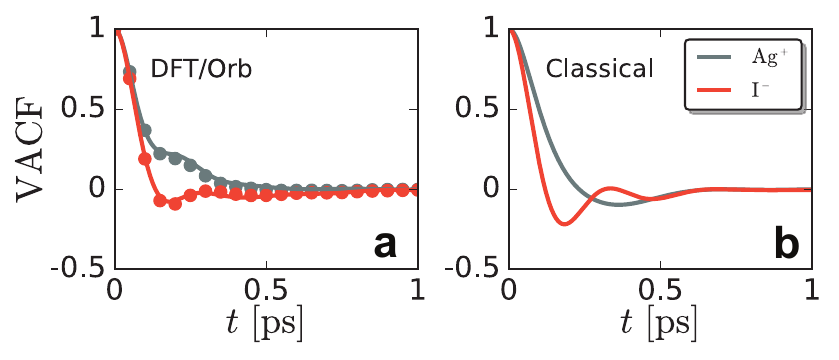}
\end{center}
\caption
{Velocity autocorrelation functions (VACFs) for Ag$^+$ and I$^-$ ions obtained from (a) DFT (points) and Orb (lines), and (b) FF models.
}
\label{fig:vacf}
\end{figure}

\subsection{Dynamics}

We use velocity autocorrelation functions (VACFs) to quantify the temporal correlations of particle motion. 
The VACF is given by
\begin{equation}
C_v(t) = \frac{\avg{\mathbf{v}(t)\cdot \mathbf{v}(0)}}{\avg{v^2(0)}},
\end{equation}
where $\mathbf{v}(t)$ is the velocity of an ion at time $t$ and $\avg{\cdots}$ indicates an ensemble average.
The DFT and Orb models yield nearly identical VACFs for both ionic species, suggesting that the Orb model accurately captures the dynamics of molten AgI, Fig.~\ref{fig:vacf}.
The diffusion coefficients obtained from integration of the unnormalized VACF, $D=\frac{1}{3}\int_0^\infty \avg{\mathbf{v}(t)\cdot \mathbf{v}(0)}dt$, indicate significantly faster diffusion of Ag$^+$ ions than I$^-$ ions, Table~\ref{tab:diffusion}.
The small silver ions diffuse significantly quicker than the slow-moving iodide ions. 
%

\begin{table}[h]
\caption{Diffusion coefficients (in units of 10$^{-5}$~cm$^2$/s) of Ag$^+$ and I$^-$ ions in molten AgI at 1600~K obtained from DFT, Orb, and FF models using two different approaches: from integration of the velocity autocorrelation function (VACF) and from inversion of the static friction coefficient, $\gamma$, as computed from integration of the memory function.}
\label{tab:diffusion}
\begin{tabular}{c|c|c|c|c}
\multirow{2}{*}{}  & \multicolumn{2}{c|}{\textbf{VACF}} & \multicolumn{2}{c}{\textbf{Friction}} \\ \cline{2-5} 
\multirow{2}{*}{}  &  \textbf{Ag$^+$}   & \textbf{I$^-$}  &  \textbf{Ag$^+$}     & \textbf{I$^-$}    \\ \hline
\textbf{DFT}        & 11.6 & 4.2 & 13.6 & 4.7 \\ 
\textbf{Orb}        & 12.6 & 4.5 & 13.8 & 4.7 \\ 
\textbf{Classical}  & 8.0 & 2.9 &  8.5 & 2.9 \\ 
\end{tabular}
\end{table}

%
The Orb model closely reproduces the diffusion coefficients obtained from DFT simulations for both ions. 
In contrast, the FF model describes only the diffusion of iodide with good accuracy and significantly underestimates the mobility of Ag$^+$ ions.
These results suggest that polarization fluctuations affect the diffusion of the cation but has only a minor impact on anion diffusion.

The VACF of I$^-$ exhibits a pronounced negative backscattering feature, arising due to the velocity reversal upon bouncing with neighboring ions. 
However, lack of oscillations suggest that caging effect is weaker than that observed in solid AgI~\cite{Dhattarwal2024paddle,dhattarwal2025electronic}.
The classical model also captures the negative dip in the VACF of I$^-$ ions, but with a slightly stronger cage effect.
The VACF of Ag$^+$ ions agree for DFT and Orb, and exhibit a positive decay in the VACF, with a bump at approximately 0.25~ps. 
The bump in the VACF indicates that the Ag$^+$ ions experience a transient positive enhancement of the correlation in velocities, suggesting collective motion due to polarization that enhances dynamics. 
In contrast, the FF model shows a negative minimum in the Ag$^+$ VACF typical of caging, consistent with the FF model underestimating the Ag$^+$ diffusion coefficient.

\begin{figure}[tb]
\begin{center}
\includegraphics[width=0.48\textwidth]{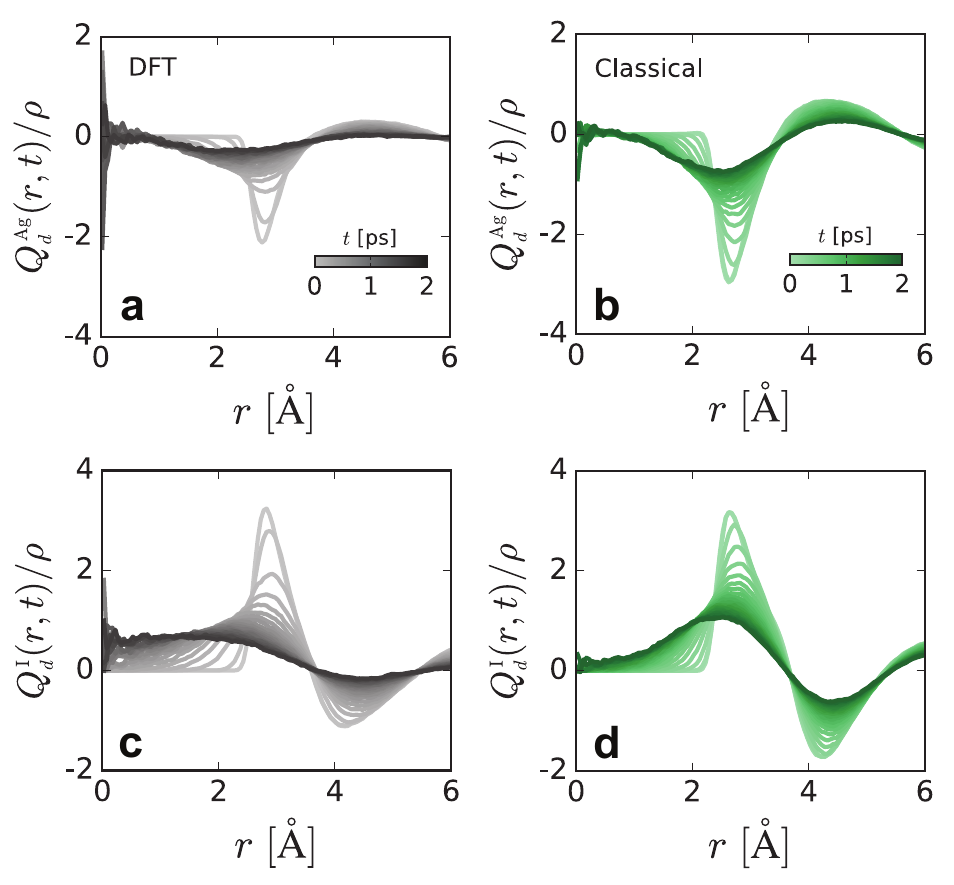}
\end{center}
\caption
{Distinct van Hove charge correlation functions for (a) and (b) Ag$^+$ and (c) and (d) I$^-$ as predicted by DFT (grey) and FF models (green).
}
\label{fig:Qd}
\end{figure}

The differences in Ag$^+$ and I$^-$ dynamics can be further characterized through space-time correlation functions,
like the distinct van Hove correlation functions, 
\begin{equation}
G_d^{X-Y}(r,t)=\avg{ \frac{1}{N} \sum_{i\in \curly{X}} \sum_{j\in \curly{Y}} \delta \para{ \len{\rb_i(0)-\rb_j(t)} } },
\end{equation}
which quantifies space-time correlations between the initial position $\rb_i(0)$ of an ion of type $X$ with all other ions of type $Y$ in the system. 
Note that at $t=0$, the distinct van Hove correlation function is equal to the $g(r)$ for correlations between species $X$ and $Y$. 
In order to simultaneously analyze contributions to space-time correlations from cations and anions, we define a distinct van Hove charge correlation function for each ion,
\begin{equation}
Q_d^{{\rm Ag}}(r,t) = q G_d^{{\rm Ag-Ag}}(r,t) - q G_d^{{\rm Ag-I}}(r,t)
\end{equation}
and 
\begin{equation}
Q_d^{{\rm I}}(r,t) = q G_d^{{\rm Ag-I}}(r,t) - q G_d^{{\rm I-I}}(r,t),
\end{equation}
where we used charges with magnitude $q=1$ to enable comparison between the Orb and FF results with a similar weighting.
The distinct van Hove charge correlation functions illustrate how the charge correlations in the system restructure as the central ion moves away from its initial location.
The cationic $Q_d^{{\rm Ag}}(r,t)$ is dominated by the first, negative peak, corresponding to the first shell of anions around the cation. 
The time-dependence of $Q_d^{{\rm Ag}}(r,t)$ from DFT suggests that the predominant structural changes are due to anions; the cation peak is small and smoothly changes with time, Fig.~\ref{fig:Qd}a. 
Near the origin, $Q_d^{{\rm Ag}}(r,t)$ remains approximately equal to zero while the peaks lose their structure, further suggesting smooth diffusive dynamics of the cation. 
The first negative peak in $Q_d^{{\rm Ag}}(r,t)$ from the FF model is shallower, as it is completely dominated by the anionic contribution. 
Here, the cation-cation interactions have higher contributions to the second peak in $Q_d^{{\rm Ag}}(r,t)$, which is more pronounced in the FF model than in DFT.
%
%

%
In contrast, the anionic $Q_d^{{\rm I}}(r,t)$ has significant structuring in both the first (positive) and second (negative) peaks, Fig.~\ref{fig:Qd}b.
The charge correlations are consistent with a more ordered coordination structure that rearranges on longer timescales.
As the peaks in $Q_d^{{\rm I}}(r,t)$ smear out over time, the value of $Q_d^{{\rm I}}(r,t)$ near the origin becomes increasingly positive. 
The positive peak at small $r$ suggests that the iodide initially at the origin is replaced by a cation as time is increased,
consistent with van Hove analyses in other ionic systems~\cite{balasubramanian1995molecular,Bala2020}.
This reflects a strong correlation between the dynamics of iodide and its silver cation coordination shell. 
Electronic polarization fluctuations effectively make Ag-I pairs more stable, such that the silver ions that coordinate an iodide replace the anion as it diffuses. 
This creates the positive peak in $Q_d^{{\rm I}}(r,t)$ at small $r$, as predicted by the Orb model. 
In contrast, cations are less tightly bound to anions in the FF model, such that $Q_d^{{\rm I}}(r,t)$ remains near zero at small $r$. 
Despite the increased strength of Ag-I interactions in the DFT and Orb models, Ag$^+$ diffuses significantly faster than in the FF model, suggesting that this pairing effect increases diffusion. 
This increased cation diffusion through strengthened Ag-I interactions is consistent with the electronic paddle-wheel effect contributing to diffusion in the liquid state; 
strong Ag-I interactions accompanied rotation of the iodide electron density facilitates Ag$^+$ translational diffusion.

The analysis of the distinct van Hove charge correlation functions therefore indicates an asymmetry between the collective dynamics of cations and anions. 
The cation dynamics are dependent on interactions with the anions, including their local electronic fluctuations through electronic rotation-cation translation coupling; similar electronic-rotation nuclear translation coupling has been identified in ionic solids~\cite{RCR_PRL_2020,RCR_APL_2020,Colin2024,hylton2025octahedral,Dhattarwal2024paddle,dhattarwal2025electronic,lian2024stereoactive,jamroz2025role}.
In contrast, the diffusion of an anion is not dictated by its dynamic cation coordination shell. 
Because the ions in that coordination shell are essentially nonpolarizable, iodide dynamics do not require an accurate description of electronic fluctuations to describe the interactions controlling their dynamics, and the resulting anion diffusion is well-described by the FF model. 

\subsection{MLWFC rotational TCF and cage correlation functions}
In superionic AgI, the hopping of Ag$^+$ from one coordination environment to another involves rotational motion of the electron density on surrounding I$^-$ ions, termed electronic paddle-wheels. 
As a result, time correlation functions that quantify the rotational motion of the anionic electron density and the hopping motion of the cations decay on the same timescale. 
Diffusion in the liquid can involve different processes than that in the solid state. 
In a liquid, cation diffusion does not require a change in coordination shell; the cation and its coordination shell could diffuse together. 
In addition, changes in the coordination shell can occur through ion pair dissociation without involving rotational motion of the anion electron density.
%

\begin{figure}[h]
\begin{center}
\includegraphics[width=0.35\textwidth]{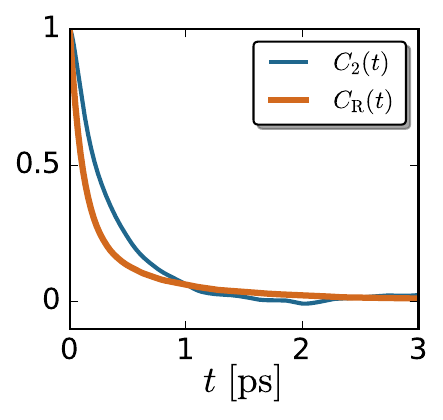}
\end{center}
\caption
{Tetrahedral rotor function time correlation functions (TCFs), $C_2(t)$, and cage residence correlation function, $C_R(t)$,
characterizing I$^-$ lone pair rotations and change in the Ag$^+$ solvation shell, respectively.
}
\label{fig:rot}
\end{figure}

%
To evaluate rotational motion of I$^-$ electron density, we represent the electron density using the centers of maximally localized Wannier functions (MLWFCs) and monitor the orientation of the unit composed of the iodide and its four MLWFCs~\cite{Dhattarwal2024paddle}. 
We monitor changes in I$^-$-MLWFC orientation using tetrahedral rotor functions~\cite{Klein1983,RCR_PRL_2020,Dhattarwal2024paddle}, $\Mb_\lambda$, of order $l=3$, such that $\lambda$ labels the $(2l+1)$ functions for each $l$, and we focus on $\lambda=2$.
The specific rotor function of interest is
\begin{equation}
\Mb_2(t) = \frac{3\sqrt{5}}{40} \sum_{i=1}^4 \para{5x_i^3 - 3x_i r_i^2},
\end{equation}
where $\rb_i=(x_i,y_i,z_i)$ is a unit vector along one of the four I-MLWFC bonds ($i$), and $r_i=\len{\rb_i}$. 
We then compute the time correlation function of this rotor function to quantify MLWFC rotations,
\begin{equation}
C_2(t) = \frac{ \avg{\Mb_2(0)\Mb_2(t)} }{ \avg{\Mb_2^2(0)}}.
\end{equation}
We quantify changes in ion coordination shell through a cage correlation function~\cite{rabani1997calculating,Dhattarwal2024paddle,dhattarwal2025electronic}.
We quantify only whether the coordination cage changes, and not by how much the cage changes, through the use of indicator functions,
\begin{equation}
h_i^{\rm out}(0,t)=\Theta\para{1-\xi^{\rm out}_i(0,t)}
\end{equation}
and
\begin{equation}
h_i^{\rm in}(0,t)=\Theta\para{1-\xi^{\rm in}_i(0,t)},
\end{equation}
where $\Theta(x)$ is the Heaviside step function,
and the functions $\xi^{\rm in}_i(0,t)$ and $\xi^{\rm out}_i(0,t)$ are the fractions of neighbors that have respectively entered or left the cage of particle $i$ between $0$ and $t$, defined by
\begin{equation}
\xi^{\rm in}_i(0,t) = \frac{ l_i(0)\cdot l_i(t)}{l_i(t)\cdot l_i(t)},
\end{equation}
and
\begin{equation}
\xi^{\rm out}_i(0,t) = \frac{ l_i(0)\cdot l_i(t)}{l_i(0)\cdot l_i(0)},
\end{equation}
where $l_i(t)$ is the neighbor list of particle $i$ at time $t$.
If the number of neighbors in an ion's cage increases between $0$ and $t$, $h_i^{\rm out}(0,t)=0$ and $h_i^{\rm in}(0,t)=1$.
If the number of neighbors in an ion's cage decreases between $0$ and $t$, $h_i^{\rm out}(0,t)=1$ and $h_i^{\rm in}(0,t)=0$.
And, if the number of neighbors in an ion's cage are the same at $0$ and $t$, then $h_i^{\rm out}(0,t)=1$ and $h_i^{\rm in}(0,t)=1$.
The indicator functions are used to compute the cage residence time correlation function,
\begin{equation}
C_{\rm R}(t) = \avg{h_i^{\rm out}(0,t)h_i^{\rm in}(0,t)},
\end{equation}
such that $C_{\rm R}(0)=1$ and $\lim_{t\rightarrow\infty}C_{\rm R}(t)=0$ for a particle that changes its coordination cage throughout the diffusion process, as expected for particles in liquids. 
While $C_2(t)$ and $C_{\rm R}(t)$ overlap in superionic AgI, the decay of $C_2(t)$ is slower than $C_{\rm R}(t)$ in liquid AgI, Fig.~\ref{fig:rot}.
The faster decay of the cage correlation function, $C_{\rm R}(t)$, indicates that the coordination environment of Ag$^+$ changes faster than the I$^-$ electron density rotates, albeit only a fraction of a picosecond faster. 
The fast change of coordination environment is due to ion pair dissociation, which need not be accompanied by rotations of the iodide electron density. 
Therefore, Ag$^+$ diffusion is expected to involve a mechanism in which orientational fluctuations of I$^-$ electron density couple to cation translations, but also another mechanism that involves purely translational diffusion and change of coordination shells through ion unpairing.

\subsection{Force fluctuations and memory functions}

\begin{figure}[tb]
\begin{center}
\includegraphics[width=0.48\textwidth]{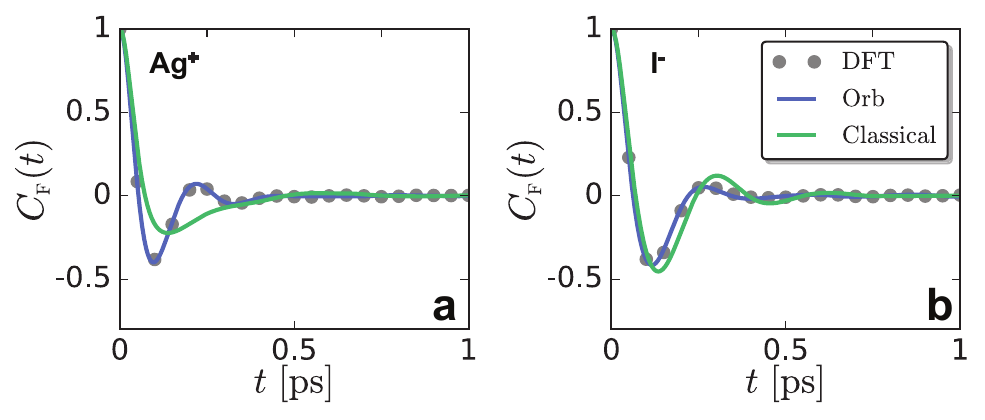}
\end{center}
\caption
{Force autocorrelation functions (FACFs) for (a) Ag$^+$ and (b) I$^-$ ions obtained from DFT, Orb, and FF models.
}
\label{fig:facf}
\end{figure}

If fluctuations in the iodide electron density are indeed impacting ionic dynamics in liquid AgI, effects of electronic fluctuations should manifest in the dynamic forces felt by diffusing ions, as well as the time-dependent friction or memory function felt by ions. 
We first focus on quantifying force fluctuations through the force autocorrelation function (FACF), Fig.~\ref{fig:facf}, defined as
\begin{equation}
C_{\rm F}(t) = \frac{ \avg{\mathbf{F}(t)\cdot \mathbf{F}(0)} } { \avg{F^2}},
\end{equation}
where $\mathbf{F}(t)$ is the force on the ion of interest at time $t$ with magnitude $F=\len{\mathbf{F}}$.
The FACFs of Ag$^+$ from DFT and the Orb model are nearly identical and are qualitatively different than that produced by the FF model.
The DFT/Orb FACF exhibits a fast, oscillatory decay with a narrow minimum at short times, while the FF model produces a FACF with a single minimum followed by a smooth decay to zero. 
Because there is no backscattering observed for the Ag$^+$ ions, as discussed for the VACF,
these high frequency force fluctuations help  Ag$^+$ ions break out of their I$^-$ cages. 
The high frequency force fluctuations in the DFT/Orb models originate in the coupling between rotational electronic fluctuations of iodide anions and translational motion of silver cations. 
%

%
The iodide FACFs are similar for all three models, as one might expect from the similar level of agreement between the iodide VACFs; the FACF is the second derivative of the VACF. 
The force fluctuations on I$^-$ ions are most likely due to collisions with neighboring silver cations.
Because the silver ions are essentially nonpolarizable, the FF model can capture the relevant interactions without the need for electronic fluctuations. 
Within the framework of the generalized Langevin equation, the time-dependent friction or memory function, $\Gamma (t)$, connects instantaneous forces on an ion to its long-time transport. 
We compute the memory function for both cation and anion by discretizing and iteratively solving the integral equation~\cite{harp1970time,berne1970calculation,boon1991molecular,daldrop2019mass}
\begin{equation}
m \frac{d C_v(t)}{dt} = -\int_0^t dt' \Gamma(t-t') C_v(t').
\end{equation}
The static friction coefficient on an ion, $\gamma$, is given by the integral of the memory function and the diffusion coefficient through, 
\begin{equation}
\gamma=\int_0^\infty \Gamma(t)dt = \frac{\kT}{D}. 
\end{equation}
Integration of the memory function produces diffusion coefficients in agreement with those obtained from the VACFs and the MDSs, Table~\ref{tab:diffusion}, suggesting that the computed memory functions provide an accurate description of the relevant ionic dynamics. 
%

\begin{figure}[tb]
\begin{center}
\includegraphics[width=0.48\textwidth]{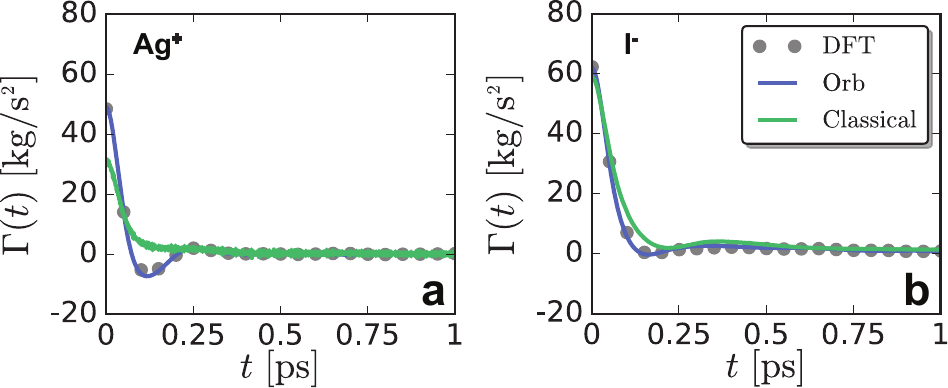}
\end{center}
\caption
{Memory function, $\Gamma(t)$, for (a) Ag$^+$ and (b) I$^-$ ions obtained from DFT, Orb, and FF models.
}
\label{fig:gamma}
\end{figure}

\begin{figure}[tb]
\begin{center}
\includegraphics[width=0.48\textwidth]{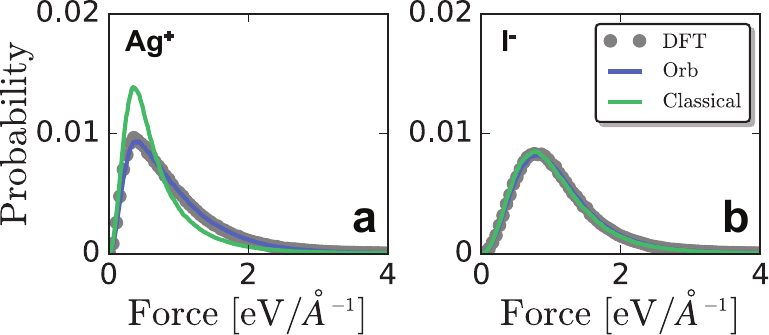}
\end{center}
\caption
{Probability distribution of the magnitude of the force, $P(F)$, on (a) a cation and (b) an anion in molten AgI.}
\label{fig:pf}
\end{figure}

%
The DFT/Orb memory functions for Ag$^+$ exhibit an oscillatory decay, Fig.~\ref{fig:gamma}, while the FF model produces a $\Gamma(t)$ with a relatively smooth, monotonic decay with no oscillatory response
The oscillatory decay in the DFT/Orb $\Gamma(t)$ has been associated with rotational motion of local electron density of the surrounding anions in solid-state AgI~\cite{dhattarwal2025electronic} and here we can similarly attribute these oscillations to local electron density fluctuations and the many-body forces they create, which are neglected in the FF model. 
In the DFT and Orb models, the negative regions in $\Gamma(t)$ effectively lower $\gamma$, leading to rapid diffusion of silver ions. 
In contrast, the FF $\Gamma(t)$ indicates that the memory is longer lived, and its smooth positive decay results in a larger $\gamma$ than in DFT and Orb, and consequently a smaller diffusion coefficient for Ag$^+$. 
We find that electronic fluctuations are less important for the iodide memory function. 
The $\Gamma(t)$ produced by the FF model is in close quantitative agreement with the DFT and Orb models, although DFT and Orb memory functions decay slightly faster. 
The similar decays of $\Gamma(t)$ for all models support the finding that iodide dynamics in the molten salt do not require an explicit description of electronic fluctuations; averaged pairwise interactions present in the FF model provide a reasonably accurate description. 
The above results regarding electronic fluctuation effects on $C_{\rm F}(t)$ and $\Gamma(t)$ can be further understood by simply examining the distribution of forces on each ion, $P(F)$, Fig.~\ref{fig:pf}. 
The distribution of forces is approximately of Maxwell-Boltzmann form near the peak, but exhibit long, approximately exponential tails at large values of $F$, which are a manifestation of strong short-range interactions. 
The distribution of forces on a cation is nearly identical for the DFT and Orb models, but the FF model is more narrow and has smaller large $F$ tails. 
This is consistent with the DFT and Orb models including strong short-range interactions that arise from electronic fluctuations of the iodide akin to electronic paddle-wheels in solid-state ion conductors. 
The FF model does not contain a description of such effects and produces a narrower distribution. 
In contrast, all three models produce essentially the same force distribution on an anion, such that electronic fluctuations do not impact forces on iodides in molten AgI. 
%

\section{Conclusions}

We have examined the role of electronic fluctuations in determining the ionic dynamics of molten AgI with molecular simulations employing ab initio DFT, machine learning-based neural network (Orb), and classical force field models of interatomic interactions. 
We find that electronic fluctuations are critically important for cation dynamics but not anion dynamics. 
This somewhat counterintuitive finding is explained by the importance of iodide in determining Ag$^+$ dynamics. 
Local electronic fluctuations of iodide ions in the coordination shell of a silver cation lead to complicated, time-dependent many-body interactions that qualitatively alter the resulting time-dependent friction felt by the cation. 
As a result, the qualitative form of quantities like force fluctuations and the velocity autocorrelation function of the silver ions are significantly different than those produced by a classical FF model that lacks electronic fluctuations and their resulting many-body forces. 
These qualitative changes lead to enhanced cation diffusion compared to the FF model. 
In contrast, iodide dynamics are not nearly as dependent on local electronic fluctuations, because they are determined mainly by interactions with essentially nonpolarizable cations. 
In comparing different models to study electronic fluctuations in molten salts, we found that the universal Orb model is in quantitative agreement with DFT-based AIMD simulations for all structural and dynamic quantities analyzed here. 
While similar agreement was also found for the NequIP model explicitly trained on DFT calculations for many configurations of solid and liquid AgI~\cite{dhattarwal2025electronic}, the Orb model was not trained on the liquid AgI system studied here and was used without any refinement of the model.
The accuracy of the Orb model suggests that some universal machine learning force field models can quantitatively model dynamics in molten salts without further refinement, avoiding the costly generation of training data and ultimately speeding up quantitative modeling of concentrated electrolytes. 
We expect that the use of accurate machine learning force fields can enable accurate simulations of electrolyte-electrode interfaces that include the many-body effects of electronic polarization and their response to external fields, which may impact the structure, dynamics, and even chemistry in the electric double layer~\cite{Gao2022SCFNN,Dhattarwal2023SCFNN,Dhattarwal2024hybrid,fedorov2014ionic,zhang2024molecular,zhang2025tuning}.
Moreover, because polarization effects are short-ranged, the results of simulations with these machine learning force fields  could be combined with advances in classical density functional theory to learn the necessary free energy functional and efficiently and accurately predict the structure and dynamics of ionic fluids~\cite{bui2025learning}.

Our findings extend a growing body of work showing that induced electronic polarization and many-body interactions qualitatively alter structural and transport properties of molten salts.
Previous studies using polarizable classical models and ab initio simulations have demonstrated that explicit treatment of anion polarization improves agreement with structural probes and transport coefficients in a variety of halide and chloride melts~\cite{Madden1993MgCl2,Madden1996quadrupole,Salanne2011polarization,aguado2003multipoles,madden1996covalent,Trullas2008polarization}
Our results build on this previous body work and additionally show that the same physics also manifests in time-dependent force fluctuations and memory kernels that control cation transport.
Moreover, recent joint experimental and simulation investigations involving X-ray and neutron scattering, Raman spectroscopy, and AIMD simulations, have emphasized the importance of many-body correlations in molten salts and provided direct experimental motivation for including polarization in models of transport~\cite{roy2021unraveling,roy2020structure,roy2021holistic,maltsev2024transient}, and our results are consistent with these conclusions. 
The oscillatory memory kernels and the strong coupling between I$^-$ electronic reorientation and Ag$^+$ motion observed here mirror the electronic paddle-wheel mechanisms identified in superionic $\alpha$-AgI~\cite{Dhattarwal2024paddle,dhattarwal2025electronic} and other solid-state electrolytes~\cite{lian2024stereoactive,jamroz2025role}. 
This correspondence suggests a conceptual continuity between polarization-mediated transport in crystalline superionic conductors and polarization-driven enhancements of cation mobility in the melt.
Understanding how directional electronic fluctuations facilitate cation motion in both phases suggests strategies to enhance ionic conductivity through anion engineering and tuning local polarizability, including through the introduction of interfaces. 
Our findings on molten AgI combined with the previous identification of solid-state electronic paddle-wheel results indicate that directional electronic fluctuations are a unifying motif controlling cation mobility in both crystalline and liquid ionic conductors.
%

\section{Acknowledgements}
This work was supported by the U.S. Army DEVCOM ARL Army Research Office (ARO) Energy Sciences Competency, Advanced Energy Materials Program award \# W911NF-24-1-0200. The views and conclusions contained in this document are those of the authors and should not be interpreted as representing the official policies, either expressed or implied, of the U.S. Army or the U.S. Government.
We acknowledge the Office of Advanced Research Computing (OARC) at Rutgers, The State University of New Jersey for providing access to the Amarel cluster and associated research computing resources that have contributed to the results reported here.
We thank Tim Duignan from Orbital Materials for help with setting up the Orb model. 


\bibliography{AgI-references}

\begin{thebibliography}{50}%
\makeatletter
\providecommand \@ifxundefined [1]{%
 \@ifx{#1\undefined}
}%
\providecommand \@ifnum [1]{%
 \ifnum #1\expandafter \@firstoftwo
 \else \expandafter \@secondoftwo
 \fi
}%
\providecommand \@ifx [1]{%
 \ifx #1\expandafter \@firstoftwo
 \else \expandafter \@secondoftwo
 \fi
}%
\providecommand \natexlab [1]{#1}%
\providecommand \enquote  [1]{``#1''}%
\providecommand \bibnamefont  [1]{#1}%
\providecommand \bibfnamefont [1]{#1}%
\providecommand \citenamefont [1]{#1}%
\providecommand \href@noop [0]{\@secondoftwo}%
\providecommand \href [0]{\begingroup \@sanitize@url \@href}%
\providecommand \@href[1]{\@@startlink{#1}\@@href}%
\providecommand \@@href[1]{\endgroup#1\@@endlink}%
\providecommand \@sanitize@url [0]{\catcode `\\12\catcode `\$12\catcode
  `\&12\catcode `\#12\catcode `\^12\catcode `\_12\catcode `\%12\relax}%
\providecommand \@@startlink[1]{}%
\providecommand \@@endlink[0]{}%
\providecommand \url  [0]{\begingroup\@sanitize@url \@url }%
\providecommand \@url [1]{\endgroup\@href {#1}{\urlprefix }}%
\providecommand \urlprefix  [0]{URL }%
\providecommand \Eprint [0]{\href }%
\providecommand \doibase [0]{http://dx.doi.org/}%
\providecommand \selectlanguage [0]{\@gobble}%
\providecommand \bibinfo  [0]{\@secondoftwo}%
\providecommand \bibfield  [0]{\@secondoftwo}%
\providecommand \translation [1]{[#1]}%
\providecommand \BibitemOpen [0]{}%
\providecommand \bibitemStop [0]{}%
\providecommand \bibitemNoStop [0]{.\EOS\space}%
\providecommand \EOS [0]{\spacefactor3000\relax}%
\providecommand \BibitemShut  [1]{\csname bibitem#1\endcsname}%
\let\auto@bib@innerbib\@empty
\bibitem [{\citenamefont {Salanne}\ and\ \citenamefont
  {Madden}(2011)}]{Salanne2011polarization}%
  \BibitemOpen
  \bibfield  {author} {\bibinfo {author} {\bibfnamefont {M.}~\bibnamefont
  {Salanne}}\ and\ \bibinfo {author} {\bibfnamefont {P.~A.}\ \bibnamefont
  {Madden}},\ }\bibfield  {title} {\enquote {\bibinfo {title} {Polarization
  effects in ionic solids and melts},}\ }\href@noop {} {\bibfield  {journal}
  {\bibinfo  {journal} {Molecular Physics}\ }\textbf {\bibinfo {volume}
  {109}},\ \bibinfo {pages} {2299--2315} (\bibinfo {year} {2011})}\BibitemShut
  {NoStop}%
\bibitem [{\citenamefont {Wilson}\ and\ \citenamefont
  {Madden}(1993)}]{Madden1993MgCl2}%
  \BibitemOpen
  \bibfield  {author} {\bibinfo {author} {\bibfnamefont {M.}~\bibnamefont
  {Wilson}}\ and\ \bibinfo {author} {\bibfnamefont {P.~A.}\ \bibnamefont
  {Madden}},\ }\bibfield  {title} {\enquote {\bibinfo {title} {Short- and
  intermediate-range order in mcl$_2$ melts: the importance of anionic
  polarization},}\ }\href@noop {} {\bibfield  {journal} {\bibinfo  {journal}
  {Journal of Physics: Condensed Matter}\ }\textbf {\bibinfo {volume} {5}},\
  \bibinfo {pages} {6833} (\bibinfo {year} {1993})}\BibitemShut {NoStop}%
\bibitem [{\citenamefont {Porter}\ \emph {et~al.}(2022)\citenamefont {Porter},
  \citenamefont {Vaka}, \citenamefont {Steenblik},\ and\ \citenamefont
  {Della~Corte}}]{Porter2022}%
  \BibitemOpen
  \bibfield  {author} {\bibinfo {author} {\bibfnamefont {T.}~\bibnamefont
  {Porter}}, \bibinfo {author} {\bibfnamefont {M.~M.}\ \bibnamefont {Vaka}},
  \bibinfo {author} {\bibfnamefont {P.}~\bibnamefont {Steenblik}}, \ and\
  \bibinfo {author} {\bibfnamefont {D.}~\bibnamefont {Della~Corte}},\
  }\bibfield  {title} {\enquote {\bibinfo {title} {Computational methods to
  simulate molten salt thermophysical properties},}\ }\href@noop {} {\bibfield
  {journal} {\bibinfo  {journal} {Communications Chemistry}\ }\textbf {\bibinfo
  {volume} {5}},\ \bibinfo {pages} {69} (\bibinfo {year} {2022})}\BibitemShut
  {NoStop}%
\bibitem [{\citenamefont {Borodin}(2009)}]{borodin2009polarizable}%
  \BibitemOpen
  \bibfield  {author} {\bibinfo {author} {\bibfnamefont {O.}~\bibnamefont
  {Borodin}},\ }\bibfield  {title} {\enquote {\bibinfo {title} {Polarizable
  force field development and molecular dynamics simulations of ionic
  liquids},}\ }\href@noop {} {\bibfield  {journal} {\bibinfo  {journal} {The
  Journal of Physical Chemistry B}\ }\textbf {\bibinfo {volume} {113}},\
  \bibinfo {pages} {11463--11478} (\bibinfo {year} {2009})}\BibitemShut
  {NoStop}%
\bibitem [{\citenamefont {Bedrov}\ \emph {et~al.}(2019)\citenamefont {Bedrov},
  \citenamefont {Piquemal}, \citenamefont {Borodin}, \citenamefont {MacKerell},
  \citenamefont {Roux},\ and\ \citenamefont {Schr{\"o}der}}]{Bedrov2019}%
  \BibitemOpen
  \bibfield  {author} {\bibinfo {author} {\bibfnamefont {D.}~\bibnamefont
  {Bedrov}}, \bibinfo {author} {\bibfnamefont {J.-P.}\ \bibnamefont
  {Piquemal}}, \bibinfo {author} {\bibfnamefont {O.}~\bibnamefont {Borodin}},
  \bibinfo {author} {\bibfnamefont {A.~D.~J.}\ \bibnamefont {MacKerell}},
  \bibinfo {author} {\bibfnamefont {B.}~\bibnamefont {Roux}}, \ and\ \bibinfo
  {author} {\bibfnamefont {C.}~\bibnamefont {Schr{\"o}der}},\ }\bibfield
  {title} {\enquote {\bibinfo {title} {Molecular dynamics simulations of ionic
  liquids and electrolytes using polarizable force fields},}\ }\href@noop {}
  {\bibfield  {journal} {\bibinfo  {journal} {Chem. Rev.}\ }\textbf {\bibinfo
  {volume} {119}},\ \bibinfo {pages} {7940--7995} (\bibinfo {year}
  {2019})}\BibitemShut {NoStop}%
\bibitem [{\citenamefont {Bedrov}\ \emph {et~al.}(2010)\citenamefont {Bedrov},
  \citenamefont {Borodin}, \citenamefont {Li},\ and\ \citenamefont
  {Smith}}]{bedrov2010influence}%
  \BibitemOpen
  \bibfield  {author} {\bibinfo {author} {\bibfnamefont {D.}~\bibnamefont
  {Bedrov}}, \bibinfo {author} {\bibfnamefont {O.}~\bibnamefont {Borodin}},
  \bibinfo {author} {\bibfnamefont {Z.}~\bibnamefont {Li}}, \ and\ \bibinfo
  {author} {\bibfnamefont {G.~D.}\ \bibnamefont {Smith}},\ }\bibfield  {title}
  {\enquote {\bibinfo {title} {Influence of polarization on structural,
  thermodynamic, and dynamic properties of ionic liquids obtained from
  molecular dynamics simulations},}\ }\href@noop {} {\bibfield  {journal}
  {\bibinfo  {journal} {The Journal of Physical Chemistry B}\ }\textbf
  {\bibinfo {volume} {114}},\ \bibinfo {pages} {4984--4997} (\bibinfo {year}
  {2010})}\BibitemShut {NoStop}%
\bibitem [{\citenamefont {Wilson}, \citenamefont {Madden},\ and\ \citenamefont
  {Costa-Cabral}(1996)}]{Madden1996quadrupole}%
  \BibitemOpen
  \bibfield  {author} {\bibinfo {author} {\bibfnamefont {M.}~\bibnamefont
  {Wilson}}, \bibinfo {author} {\bibfnamefont {P.~A.}\ \bibnamefont {Madden}},
  \ and\ \bibinfo {author} {\bibfnamefont {B.~J.}\ \bibnamefont
  {Costa-Cabral}},\ }\bibfield  {title} {\enquote {\bibinfo {title} {Quadrupole
  polarization in simulations of ionic systems: Application to agcl},}\
  }\href@noop {} {\bibfield  {journal} {\bibinfo  {journal} {The Journal of
  Physical Chemistry}\ }\textbf {\bibinfo {volume} {100}},\ \bibinfo {pages}
  {1227--1237} (\bibinfo {year} {1996})}\BibitemShut {NoStop}%
\bibitem [{\citenamefont {Madden}\ and\ \citenamefont
  {Wilson}(1996)}]{madden1996covalent}%
  \BibitemOpen
  \bibfield  {author} {\bibinfo {author} {\bibfnamefont {P.~A.}\ \bibnamefont
  {Madden}}\ and\ \bibinfo {author} {\bibfnamefont {M.}~\bibnamefont
  {Wilson}},\ }\bibfield  {title} {\enquote {\bibinfo {title} {`covalent'
  effects in `ionic' systems},}\ }\href@noop {} {\bibfield  {journal} {\bibinfo
   {journal} {Chemical Society Reviews}\ }\textbf {\bibinfo {volume} {25}},\
  \bibinfo {pages} {339--350} (\bibinfo {year} {1996})}\BibitemShut {NoStop}%
\bibitem [{\citenamefont {Bitri\`{a}n}\ and\ \citenamefont
  {Trull\`{a}s}(2008)}]{Trullas2008polarization}%
  \BibitemOpen
  \bibfield  {author} {\bibinfo {author} {\bibfnamefont {V.}~\bibnamefont
  {Bitri\`{a}n}}\ and\ \bibinfo {author} {\bibfnamefont {J.}~\bibnamefont
  {Trull\`{a}s}},\ }\bibfield  {title} {\enquote {\bibinfo {title} {Molecular
  dynamics study of polarization effects on agi},}\ }\href@noop {} {\bibfield
  {journal} {\bibinfo  {journal} {The Journal of Physical Chemistry B}\
  }\textbf {\bibinfo {volume} {112}},\ \bibinfo {pages} {1718--1728} (\bibinfo
  {year} {2008})}\BibitemShut {NoStop}%
\bibitem [{\citenamefont {Dhattarwal}, \citenamefont {Somni},\ and\
  \citenamefont {Remsing}(2024)}]{Dhattarwal2024paddle}%
  \BibitemOpen
  \bibfield  {author} {\bibinfo {author} {\bibfnamefont {H.~S.}\ \bibnamefont
  {Dhattarwal}}, \bibinfo {author} {\bibfnamefont {R.}~\bibnamefont {Somni}}, \
  and\ \bibinfo {author} {\bibfnamefont {R.~C.}\ \bibnamefont {Remsing}},\
  }\bibfield  {title} {\enquote {\bibinfo {title} {Electronic paddle-wheels in
  a solid-state electrolyte},}\ }\href@noop {} {\bibfield  {journal} {\bibinfo
  {journal} {Nat. Commun.}\ }\textbf {\bibinfo {volume} {15}},\ \bibinfo
  {pages} {121} (\bibinfo {year} {2024})}\BibitemShut {NoStop}%
\bibitem [{\citenamefont {Vashishta}\ and\ \citenamefont
  {Rahman}(1978)}]{AgI_Vashishta}%
  \BibitemOpen
  \bibfield  {author} {\bibinfo {author} {\bibfnamefont {P.}~\bibnamefont
  {Vashishta}}\ and\ \bibinfo {author} {\bibfnamefont {A.}~\bibnamefont
  {Rahman}},\ }\bibfield  {title} {\enquote {\bibinfo {title} {Ionic motion in
  $\ensuremath{\alpha}$-agi},}\ }\href@noop {} {\bibfield  {journal} {\bibinfo
  {journal} {Phys. Rev. Lett.}\ }\textbf {\bibinfo {volume} {40}},\ \bibinfo
  {pages} {1337--1340} (\bibinfo {year} {1978})}\BibitemShut {NoStop}%
\bibitem [{\citenamefont {{Rhodes}}\ \emph {et~al.}(2025)\citenamefont
  {{Rhodes}}, \citenamefont {{Vandenhaute}}, \citenamefont {{{\v{S}}imkus}},
  \citenamefont {{Gin}}, \citenamefont {{Godwin}}, \citenamefont {{Duignan}},\
  and\ \citenamefont {{Neumann}}}]{Orb2025}%
  \BibitemOpen
  \bibfield  {author} {\bibinfo {author} {\bibfnamefont {B.}~\bibnamefont
  {{Rhodes}}}, \bibinfo {author} {\bibfnamefont {S.}~\bibnamefont
  {{Vandenhaute}}}, \bibinfo {author} {\bibfnamefont {V.}~\bibnamefont
  {{{\v{S}}imkus}}}, \bibinfo {author} {\bibfnamefont {J.}~\bibnamefont
  {{Gin}}}, \bibinfo {author} {\bibfnamefont {J.}~\bibnamefont {{Godwin}}},
  \bibinfo {author} {\bibfnamefont {T.}~\bibnamefont {{Duignan}}}, \ and\
  \bibinfo {author} {\bibfnamefont {M.}~\bibnamefont {{Neumann}}},\ }\bibfield
  {title} {\enquote {\bibinfo {title} {{Orb-v3: atomistic simulation at
  scale}},}\ }\href {\doibase 10.48550/arXiv.2504.06231} {\bibfield  {journal}
  {\bibinfo  {journal} {arXiv e-prints}\ ,\ \bibinfo {eid} {arXiv:2504.06231}}
  (\bibinfo {year} {2025})},\ \Eprint {http://arxiv.org/abs/2504.06231}
  {arXiv:2504.06231 [cond-mat.mtrl-sci]} \BibitemShut {NoStop}%
\bibitem [{\citenamefont {K\"{u}hne}\ \emph {et~al.}(2020)\citenamefont
  {K\"{u}hne}, \citenamefont {Iannuzzi}, \citenamefont {Del~Ben}, \citenamefont
  {Rybkin}, \citenamefont {Seewald}, \citenamefont {Stein}, \citenamefont
  {Laino}, \citenamefont {Khaliullin}, \citenamefont {Sch\"{u}tt},
  \citenamefont {Schiffmann}, \citenamefont {Golze}, \citenamefont {Wilhelm},
  \citenamefont {Chulkov}, \citenamefont {Bani-Hashemian}, \citenamefont
  {Weber}, \citenamefont {Bor{\v s}tnik}, \citenamefont {Taillefumier},
  \citenamefont {Jakobovits}, \citenamefont {Lazzaro}, \citenamefont {Pabst},
  \citenamefont {M\"{u}ller}, \citenamefont {Schade}, \citenamefont {Guidon},
  \citenamefont {Andermatt}, \citenamefont {Holmberg}, \citenamefont
  {Schenter}, \citenamefont {Hehn}, \citenamefont {Bussy}, \citenamefont
  {Belleflamme}, \citenamefont {Tabacchi}, \citenamefont {Gl{\"o}{\ss}},
  \citenamefont {Lass}, \citenamefont {Bethune}, \citenamefont {Mundy},
  \citenamefont {Plessl}, \citenamefont {Watkins}, \citenamefont
  {VandeVondele}, \citenamefont {Krack},\ and\ \citenamefont
  {Hutter}}]{CP2K2020}%
  \BibitemOpen
  \bibfield  {author} {\bibinfo {author} {\bibfnamefont {T.~D.}\ \bibnamefont
  {K\"{u}hne}}, \bibinfo {author} {\bibfnamefont {M.}~\bibnamefont {Iannuzzi}},
  \bibinfo {author} {\bibfnamefont {M.}~\bibnamefont {Del~Ben}}, \bibinfo
  {author} {\bibfnamefont {V.~V.}\ \bibnamefont {Rybkin}}, \bibinfo {author}
  {\bibfnamefont {P.}~\bibnamefont {Seewald}}, \bibinfo {author} {\bibfnamefont
  {F.}~\bibnamefont {Stein}}, \bibinfo {author} {\bibfnamefont
  {T.}~\bibnamefont {Laino}}, \bibinfo {author} {\bibfnamefont {R.~Z.}\
  \bibnamefont {Khaliullin}}, \bibinfo {author} {\bibfnamefont
  {O.}~\bibnamefont {Sch\"{u}tt}}, \bibinfo {author} {\bibfnamefont
  {F.}~\bibnamefont {Schiffmann}}, \bibinfo {author} {\bibfnamefont
  {D.}~\bibnamefont {Golze}}, \bibinfo {author} {\bibfnamefont
  {J.}~\bibnamefont {Wilhelm}}, \bibinfo {author} {\bibfnamefont
  {S.}~\bibnamefont {Chulkov}}, \bibinfo {author} {\bibfnamefont {M.~H.}\
  \bibnamefont {Bani-Hashemian}}, \bibinfo {author} {\bibfnamefont
  {V.}~\bibnamefont {Weber}}, \bibinfo {author} {\bibfnamefont
  {U.}~\bibnamefont {Bor{\v s}tnik}}, \bibinfo {author} {\bibfnamefont
  {M.}~\bibnamefont {Taillefumier}}, \bibinfo {author} {\bibfnamefont {A.~S.}\
  \bibnamefont {Jakobovits}}, \bibinfo {author} {\bibfnamefont
  {A.}~\bibnamefont {Lazzaro}}, \bibinfo {author} {\bibfnamefont
  {H.}~\bibnamefont {Pabst}}, \bibinfo {author} {\bibfnamefont
  {T.}~\bibnamefont {M\"{u}ller}}, \bibinfo {author} {\bibfnamefont
  {R.}~\bibnamefont {Schade}}, \bibinfo {author} {\bibfnamefont
  {M.}~\bibnamefont {Guidon}}, \bibinfo {author} {\bibfnamefont
  {S.}~\bibnamefont {Andermatt}}, \bibinfo {author} {\bibfnamefont
  {N.}~\bibnamefont {Holmberg}}, \bibinfo {author} {\bibfnamefont {G.~K.}\
  \bibnamefont {Schenter}}, \bibinfo {author} {\bibfnamefont {A.}~\bibnamefont
  {Hehn}}, \bibinfo {author} {\bibfnamefont {A.}~\bibnamefont {Bussy}},
  \bibinfo {author} {\bibfnamefont {F.}~\bibnamefont {Belleflamme}}, \bibinfo
  {author} {\bibfnamefont {G.}~\bibnamefont {Tabacchi}}, \bibinfo {author}
  {\bibfnamefont {A.}~\bibnamefont {Gl{\"o}{\ss}}}, \bibinfo {author}
  {\bibfnamefont {M.}~\bibnamefont {Lass}}, \bibinfo {author} {\bibfnamefont
  {I.}~\bibnamefont {Bethune}}, \bibinfo {author} {\bibfnamefont {C.~J.}\
  \bibnamefont {Mundy}}, \bibinfo {author} {\bibfnamefont {C.}~\bibnamefont
  {Plessl}}, \bibinfo {author} {\bibfnamefont {M.}~\bibnamefont {Watkins}},
  \bibinfo {author} {\bibfnamefont {J.}~\bibnamefont {VandeVondele}}, \bibinfo
  {author} {\bibfnamefont {M.}~\bibnamefont {Krack}}, \ and\ \bibinfo {author}
  {\bibfnamefont {J.}~\bibnamefont {Hutter}},\ }\bibfield  {title} {\enquote
  {\bibinfo {title} {{CP2K: An electronic structure and molecular dynamics
  software package - Quickstep: Efficient and accurate electronic structure
  calculations}},}\ }\href@noop {} {\bibfield  {journal} {\bibinfo  {journal}
  {J. Chem. Phys.}\ }\textbf {\bibinfo {volume} {152}},\ \bibinfo {pages}
  {194103} (\bibinfo {year} {2020})}\BibitemShut {NoStop}%
\bibitem [{\citenamefont {Perdew}, \citenamefont {Burke},\ and\ \citenamefont
  {Ernzerhof}(1996)}]{PBE1996}%
  \BibitemOpen
  \bibfield  {author} {\bibinfo {author} {\bibfnamefont {J.~P.}\ \bibnamefont
  {Perdew}}, \bibinfo {author} {\bibfnamefont {K.}~\bibnamefont {Burke}}, \
  and\ \bibinfo {author} {\bibfnamefont {M.}~\bibnamefont {Ernzerhof}},\
  }\bibfield  {title} {\enquote {\bibinfo {title} {Generalized gradient
  approximation made simple},}\ }\href@noop {} {\bibfield  {journal} {\bibinfo
  {journal} {Phys. Rev. Lett.}\ }\textbf {\bibinfo {volume} {77}},\ \bibinfo
  {pages} {3865--3868} (\bibinfo {year} {1996})}\BibitemShut {NoStop}%
\bibitem [{\citenamefont {Goedecker}, \citenamefont {Teter},\ and\
  \citenamefont {Hutter}(1996)}]{GTH1996}%
  \BibitemOpen
  \bibfield  {author} {\bibinfo {author} {\bibfnamefont {S.}~\bibnamefont
  {Goedecker}}, \bibinfo {author} {\bibfnamefont {M.}~\bibnamefont {Teter}}, \
  and\ \bibinfo {author} {\bibfnamefont {J.}~\bibnamefont {Hutter}},\
  }\bibfield  {title} {\enquote {\bibinfo {title} {Separable dual-space
  gaussian pseudopotentials},}\ }\href@noop {} {\bibfield  {journal} {\bibinfo
  {journal} {Phys. Rev. B}\ }\textbf {\bibinfo {volume} {54}},\ \bibinfo
  {pages} {1703--1710} (\bibinfo {year} {1996})}\BibitemShut {NoStop}%
\bibitem [{\citenamefont {Krack}(2005)}]{Krack2005}%
  \BibitemOpen
  \bibfield  {author} {\bibinfo {author} {\bibfnamefont {M.}~\bibnamefont
  {Krack}},\ }\bibfield  {title} {\enquote {\bibinfo {title} {Pseudopotentials
  for h to kr optimized for gradient-corrected exchange-correlation
  functionals},}\ }\href {\doibase 10.1007/s00214-005-0655-y} {\bibfield
  {journal} {\bibinfo  {journal} {Theor. Chem. Acc.}\ }\textbf {\bibinfo
  {volume} {114}},\ \bibinfo {pages} {145--152} (\bibinfo {year}
  {2005})}\BibitemShut {NoStop}%
\bibitem [{\citenamefont {Dhattarwal}\ and\ \citenamefont
  {Remsing}(2025)}]{dhattarwal2025electronic}%
  \BibitemOpen
  \bibfield  {author} {\bibinfo {author} {\bibfnamefont {H.~S.}\ \bibnamefont
  {Dhattarwal}}\ and\ \bibinfo {author} {\bibfnamefont {R.~C.}\ \bibnamefont
  {Remsing}},\ }\bibfield  {title} {\enquote {\bibinfo {title} {Electronic
  paddlewheels impact the dynamics of superionic conduction in agi},}\
  }\href@noop {} {\bibfield  {journal} {\bibinfo  {journal} {arXiv preprint
  arXiv:2504.16704}\ } (\bibinfo {year} {2025})}\BibitemShut {NoStop}%
\bibitem [{\citenamefont {Bussi}, \citenamefont {Donadio},\ and\ \citenamefont
  {Parrinello}(2007)}]{CSVR2007}%
  \BibitemOpen
  \bibfield  {author} {\bibinfo {author} {\bibfnamefont {G.}~\bibnamefont
  {Bussi}}, \bibinfo {author} {\bibfnamefont {D.}~\bibnamefont {Donadio}}, \
  and\ \bibinfo {author} {\bibfnamefont {M.}~\bibnamefont {Parrinello}},\
  }\bibfield  {title} {\enquote {\bibinfo {title} {{Canonical sampling through
  velocity rescaling}},}\ }\href@noop {} {\bibfield  {journal} {\bibinfo
  {journal} {J. Chem. Phys.}\ }\textbf {\bibinfo {volume} {126}},\ \bibinfo
  {pages} {014101} (\bibinfo {year} {2007})}\BibitemShut {NoStop}%
\bibitem [{\citenamefont {Larsen}\ \emph {et~al.}(2017)\citenamefont {Larsen},
  \citenamefont {Mortensen}, \citenamefont {Blomqvist}, \citenamefont
  {Castelli}, \citenamefont {Christensen}, \citenamefont {Du{\l}ak},
  \citenamefont {Friis}, \citenamefont {Groves}, \citenamefont {Hammer},
  \citenamefont {Hargus} \emph {et~al.}}]{ASE}%
  \BibitemOpen
  \bibfield  {author} {\bibinfo {author} {\bibfnamefont {A.~H.}\ \bibnamefont
  {Larsen}}, \bibinfo {author} {\bibfnamefont {J.~J.}\ \bibnamefont
  {Mortensen}}, \bibinfo {author} {\bibfnamefont {J.}~\bibnamefont
  {Blomqvist}}, \bibinfo {author} {\bibfnamefont {I.~E.}\ \bibnamefont
  {Castelli}}, \bibinfo {author} {\bibfnamefont {R.}~\bibnamefont
  {Christensen}}, \bibinfo {author} {\bibfnamefont {M.}~\bibnamefont
  {Du{\l}ak}}, \bibinfo {author} {\bibfnamefont {J.}~\bibnamefont {Friis}},
  \bibinfo {author} {\bibfnamefont {M.~N.}\ \bibnamefont {Groves}}, \bibinfo
  {author} {\bibfnamefont {B.}~\bibnamefont {Hammer}}, \bibinfo {author}
  {\bibfnamefont {C.}~\bibnamefont {Hargus}},  \emph {et~al.},\ }\bibfield
  {title} {\enquote {\bibinfo {title} {The atomic simulation environment---a
  python library for working with atoms},}\ }\href@noop {} {\bibfield
  {journal} {\bibinfo  {journal} {Journal of Physics: Condensed Matter}\
  }\textbf {\bibinfo {volume} {29}},\ \bibinfo {pages} {273002} (\bibinfo
  {year} {2017})}\BibitemShut {NoStop}%
\bibitem [{\citenamefont {Thompson}\ \emph {et~al.}(2022)\citenamefont
  {Thompson}, \citenamefont {Aktulga}, \citenamefont {Berger}, \citenamefont
  {Bolintineanu}, \citenamefont {Brown}, \citenamefont {Crozier}, \citenamefont
  {In't~Veld}, \citenamefont {Kohlmeyer}, \citenamefont {Moore}, \citenamefont
  {Nguyen} \emph {et~al.}}]{LAMMPS}%
  \BibitemOpen
  \bibfield  {author} {\bibinfo {author} {\bibfnamefont {A.~P.}\ \bibnamefont
  {Thompson}}, \bibinfo {author} {\bibfnamefont {H.~M.}\ \bibnamefont
  {Aktulga}}, \bibinfo {author} {\bibfnamefont {R.}~\bibnamefont {Berger}},
  \bibinfo {author} {\bibfnamefont {D.~S.}\ \bibnamefont {Bolintineanu}},
  \bibinfo {author} {\bibfnamefont {W.~M.}\ \bibnamefont {Brown}}, \bibinfo
  {author} {\bibfnamefont {P.~S.}\ \bibnamefont {Crozier}}, \bibinfo {author}
  {\bibfnamefont {P.~J.}\ \bibnamefont {In't~Veld}}, \bibinfo {author}
  {\bibfnamefont {A.}~\bibnamefont {Kohlmeyer}}, \bibinfo {author}
  {\bibfnamefont {S.~G.}\ \bibnamefont {Moore}}, \bibinfo {author}
  {\bibfnamefont {T.~D.}\ \bibnamefont {Nguyen}},  \emph {et~al.},\ }\bibfield
  {title} {\enquote {\bibinfo {title} {Lammps-a flexible simulation tool for
  particle-based materials modeling at the atomic, meso, and continuum
  scales},}\ }\href@noop {} {\bibfield  {journal} {\bibinfo  {journal}
  {Computer physics communications}\ }\textbf {\bibinfo {volume} {271}},\
  \bibinfo {pages} {108171} (\bibinfo {year} {2022})}\BibitemShut {NoStop}%
\bibitem [{\citenamefont {Nos\'{e}}(1984)}]{Nose1984}%
  \BibitemOpen
  \bibfield  {author} {\bibinfo {author} {\bibfnamefont {S.}~\bibnamefont
  {Nos\'{e}}},\ }\bibfield  {title} {\enquote {\bibinfo {title} {A unified
  formulation of the constant temperature molecular dynamics methods},}\
  }\href@noop {} {\bibfield  {journal} {\bibinfo  {journal} {J. Chem. Phys.}\
  }\textbf {\bibinfo {volume} {81}},\ \bibinfo {pages} {511--519} (\bibinfo
  {year} {1984})}\BibitemShut {NoStop}%
\bibitem [{\citenamefont {Hoover}(1985)}]{Hoover1985}%
  \BibitemOpen
  \bibfield  {author} {\bibinfo {author} {\bibfnamefont {W.~G.}\ \bibnamefont
  {Hoover}},\ }\bibfield  {title} {\enquote {\bibinfo {title} {Canonical
  dynamics: Equilibrium phase-space distributions},}\ }\href@noop {} {\bibfield
   {journal} {\bibinfo  {journal} {Phys. Rev. A}\ }\textbf {\bibinfo {volume}
  {31}},\ \bibinfo {pages} {1695--1697} (\bibinfo {year} {1985})}\BibitemShut
  {NoStop}%
\bibitem [{\citenamefont {Chahid}\ and\ \citenamefont
  {McGreevy}(1998)}]{chahid1998structure}%
  \BibitemOpen
  \bibfield  {author} {\bibinfo {author} {\bibfnamefont {A.}~\bibnamefont
  {Chahid}}\ and\ \bibinfo {author} {\bibfnamefont {R.}~\bibnamefont
  {McGreevy}},\ }\bibfield  {title} {\enquote {\bibinfo {title} {Structure and
  ionic conduction in cui: diffuse neutron scattering and rmc modelling},}\
  }\href@noop {} {\bibfield  {journal} {\bibinfo  {journal} {Journal of
  Physics: Condensed Matter}\ }\textbf {\bibinfo {volume} {10}},\ \bibinfo
  {pages} {2597} (\bibinfo {year} {1998})}\BibitemShut {NoStop}%
\bibitem [{\citenamefont {Bitri{\'a}n}\ and\ \citenamefont
  {Trull{\`a}s}(2006)}]{bitrian2006molecular}%
  \BibitemOpen
  \bibfield  {author} {\bibinfo {author} {\bibfnamefont {V.}~\bibnamefont
  {Bitri{\'a}n}}\ and\ \bibinfo {author} {\bibfnamefont {J.}~\bibnamefont
  {Trull{\`a}s}},\ }\bibfield  {title} {\enquote {\bibinfo {title} {Molecular
  dynamics study of polarizable ion models for molten agbr},}\ }\href@noop {}
  {\bibfield  {journal} {\bibinfo  {journal} {The Journal of Physical Chemistry
  B}\ }\textbf {\bibinfo {volume} {110}},\ \bibinfo {pages} {7490--7499}
  (\bibinfo {year} {2006})}\BibitemShut {NoStop}%
\bibitem [{\citenamefont {Balasubramanian}\ and\ \citenamefont
  {Rao}(1995)}]{balasubramanian1995molecular}%
  \BibitemOpen
  \bibfield  {author} {\bibinfo {author} {\bibfnamefont {S.}~\bibnamefont
  {Balasubramanian}}\ and\ \bibinfo {author} {\bibfnamefont {K.}~\bibnamefont
  {Rao}},\ }\bibfield  {title} {\enquote {\bibinfo {title} {A molecular
  dynamics study of the mixed alkali effect in silicate glasses},}\ }\href@noop
  {} {\bibfield  {journal} {\bibinfo  {journal} {Journal of non-crystalline
  solids}\ }\textbf {\bibinfo {volume} {181}},\ \bibinfo {pages} {157--174}
  (\bibinfo {year} {1995})}\BibitemShut {NoStop}%
\bibitem [{\citenamefont {Mukherji}\ \emph {et~al.}(2020)\citenamefont
  {Mukherji}, \citenamefont {Avula}, \citenamefont {Kumar},\ and\ \citenamefont
  {Balasubramanian}}]{Bala2020}%
  \BibitemOpen
  \bibfield  {author} {\bibinfo {author} {\bibfnamefont {S.}~\bibnamefont
  {Mukherji}}, \bibinfo {author} {\bibfnamefont {N.~V.~S.}\ \bibnamefont
  {Avula}}, \bibinfo {author} {\bibfnamefont {R.}~\bibnamefont {Kumar}}, \ and\
  \bibinfo {author} {\bibfnamefont {S.}~\bibnamefont {Balasubramanian}},\
  }\bibfield  {title} {\enquote {\bibinfo {title} {Hopping in high
  concentration electrolytes - long time bulk and single-particle signatures,
  free energy barriers, and structural insights},}\ }\href@noop {} {\bibfield
  {journal} {\bibinfo  {journal} {J. Phys. Chem. Lett.}\ }\textbf {\bibinfo
  {volume} {11}},\ \bibinfo {pages} {9613--9620} (\bibinfo {year}
  {2020})}\BibitemShut {NoStop}%
\bibitem [{\citenamefont {Remsing}\ and\ \citenamefont
  {Klein}(2020{\natexlab{a}})}]{RCR_PRL_2020}%
  \BibitemOpen
  \bibfield  {author} {\bibinfo {author} {\bibfnamefont {R.~C.}\ \bibnamefont
  {Remsing}}\ and\ \bibinfo {author} {\bibfnamefont {M.~L.}\ \bibnamefont
  {Klein}},\ }\bibfield  {title} {\enquote {\bibinfo {title} {Lone pair
  rotational dynamics in solids},}\ }\href@noop {} {\bibfield  {journal}
  {\bibinfo  {journal} {Phys. Rev. Lett.}\ }\textbf {\bibinfo {volume} {124}},\
  \bibinfo {pages} {066001} (\bibinfo {year} {2020}{\natexlab{a}})}\BibitemShut
  {NoStop}%
\bibitem [{\citenamefont {Remsing}\ and\ \citenamefont
  {Klein}(2020{\natexlab{b}})}]{RCR_APL_2020}%
  \BibitemOpen
  \bibfield  {author} {\bibinfo {author} {\bibfnamefont {R.~C.}\ \bibnamefont
  {Remsing}}\ and\ \bibinfo {author} {\bibfnamefont {M.~L.}\ \bibnamefont
  {Klein}},\ }\bibfield  {title} {\enquote {\bibinfo {title} {{A new
  perspective on lone pair dynamics in halide perovskites}},}\ }\href@noop {}
  {\bibfield  {journal} {\bibinfo  {journal} {APL Mater.}\ }\textbf {\bibinfo
  {volume} {8}},\ \bibinfo {pages} {050902} (\bibinfo {year}
  {2020}{\natexlab{b}})}\BibitemShut {NoStop}%
\bibitem [{\citenamefont {Hylton-Farrington}\ and\ \citenamefont
  {Remsing}(2024)}]{Colin2024}%
  \BibitemOpen
  \bibfield  {author} {\bibinfo {author} {\bibfnamefont {C.~M.}\ \bibnamefont
  {Hylton-Farrington}}\ and\ \bibinfo {author} {\bibfnamefont {R.~C.}\
  \bibnamefont {Remsing}},\ }\bibfield  {title} {\enquote {\bibinfo {title}
  {Dynamic local symmetry fluctuations of electron density in halide
  perovskites},}\ }\href@noop {} {\bibfield  {journal} {\bibinfo  {journal}
  {Chem. Mater.}\ }\textbf {\bibinfo {volume} {36}},\ \bibinfo {pages}
  {9442--9459} (\bibinfo {year} {2024})}\BibitemShut {NoStop}%
\bibitem [{\citenamefont {Hylton-Farrington}\ and\ \citenamefont
  {Remsing}(2025)}]{hylton2025octahedral}%
  \BibitemOpen
  \bibfield  {author} {\bibinfo {author} {\bibfnamefont {C.~M.}\ \bibnamefont
  {Hylton-Farrington}}\ and\ \bibinfo {author} {\bibfnamefont {R.~C.}\
  \bibnamefont {Remsing}},\ }\bibfield  {title} {\enquote {\bibinfo {title}
  {Octahedral tilting and b-site off-centering in halide perovskites are not
  coupled},}\ }\href@noop {} {\bibfield  {journal} {\bibinfo  {journal} {arXiv
  preprint arXiv:2508.15607}\ } (\bibinfo {year} {2025})}\BibitemShut {NoStop}%
\bibitem [{\citenamefont {Lian}, \citenamefont {Dambournet},\ and\
  \citenamefont {Salanne}(2024)}]{lian2024stereoactive}%
  \BibitemOpen
  \bibfield  {author} {\bibinfo {author} {\bibfnamefont {X.}~\bibnamefont
  {Lian}}, \bibinfo {author} {\bibfnamefont {D.}~\bibnamefont {Dambournet}}, \
  and\ \bibinfo {author} {\bibfnamefont {M.}~\bibnamefont {Salanne}},\
  }\bibfield  {title} {\enquote {\bibinfo {title} {Stereoactive electron lone
  pairs facilitate fluoride ion diffusion in tetragonal basnf4},}\ }\href@noop
  {} {\bibfield  {journal} {\bibinfo  {journal} {Chemistry of Materials}\
  }\textbf {\bibinfo {volume} {37}},\ \bibinfo {pages} {378--386} (\bibinfo
  {year} {2024})}\BibitemShut {NoStop}%
\bibitem [{\citenamefont {Jamroz}\ \emph {et~al.}(2025)\citenamefont {Jamroz},
  \citenamefont {Krynski}, \citenamefont {Malys}, \citenamefont {Krok},
  \citenamefont {Kyriacou}, \citenamefont {Ahmed}, \citenamefont {Abrahams},\
  and\ \citenamefont {Wrobel}}]{jamroz2025role}%
  \BibitemOpen
  \bibfield  {author} {\bibinfo {author} {\bibfnamefont {J.}~\bibnamefont
  {Jamroz}}, \bibinfo {author} {\bibfnamefont {M.}~\bibnamefont {Krynski}},
  \bibinfo {author} {\bibfnamefont {M.}~\bibnamefont {Malys}}, \bibinfo
  {author} {\bibfnamefont {F.}~\bibnamefont {Krok}}, \bibinfo {author}
  {\bibfnamefont {A.}~\bibnamefont {Kyriacou}}, \bibinfo {author}
  {\bibfnamefont {S.~J.}\ \bibnamefont {Ahmed}}, \bibinfo {author}
  {\bibfnamefont {I.}~\bibnamefont {Abrahams}}, \ and\ \bibinfo {author}
  {\bibfnamefont {W.}~\bibnamefont {Wrobel}},\ }\bibfield  {title} {\enquote
  {\bibinfo {title} {The role of lone pairs in fast oxide ion conduction in
  rhombohedral bi0. 8pr0. 2o1. 5},}\ }\href@noop {} {\bibfield  {journal}
  {\bibinfo  {journal} {Chemistry of Materials}\ } (\bibinfo {year}
  {2025})}\BibitemShut {NoStop}%
\bibitem [{\citenamefont {Klein}, \citenamefont {McDonald},\ and\ \citenamefont
  {Ozaki}(1983)}]{Klein1983}%
  \BibitemOpen
  \bibfield  {author} {\bibinfo {author} {\bibfnamefont {M.~L.}\ \bibnamefont
  {Klein}}, \bibinfo {author} {\bibfnamefont {I.~R.}\ \bibnamefont {McDonald}},
  \ and\ \bibinfo {author} {\bibfnamefont {Y.}~\bibnamefont {Ozaki}},\
  }\bibfield  {title} {\enquote {\bibinfo {title} {{Orientational order in
  ionic crystals containing tetrahedral ions}},}\ }\href@noop {} {\bibfield
  {journal} {\bibinfo  {journal} {J. Chem. Phys.}\ }\textbf {\bibinfo {volume}
  {79}},\ \bibinfo {pages} {5579--5587} (\bibinfo {year} {1983})}\BibitemShut
  {NoStop}%
\bibitem [{\citenamefont {Rabani}, \citenamefont {Gezelter},\ and\
  \citenamefont {Berne}(1997)}]{rabani1997calculating}%
  \BibitemOpen
  \bibfield  {author} {\bibinfo {author} {\bibfnamefont {E.}~\bibnamefont
  {Rabani}}, \bibinfo {author} {\bibfnamefont {J.~D.}\ \bibnamefont
  {Gezelter}}, \ and\ \bibinfo {author} {\bibfnamefont {B.}~\bibnamefont
  {Berne}},\ }\bibfield  {title} {\enquote {\bibinfo {title} {Calculating the
  hopping rate for self-diffusion on rough potential energy surfaces: Cage
  correlations},}\ }\href@noop {} {\bibfield  {journal} {\bibinfo  {journal}
  {J. Chem. Phys.}\ }\textbf {\bibinfo {volume} {107}},\ \bibinfo {pages}
  {6867--6876} (\bibinfo {year} {1997})}\BibitemShut {NoStop}%
\bibitem [{\citenamefont {Harp}\ and\ \citenamefont
  {Berne}(1970)}]{harp1970time}%
  \BibitemOpen
  \bibfield  {author} {\bibinfo {author} {\bibfnamefont {G.}~\bibnamefont
  {Harp}}\ and\ \bibinfo {author} {\bibfnamefont {B.}~\bibnamefont {Berne}},\
  }\bibfield  {title} {\enquote {\bibinfo {title} {Time-correlation functions,
  memory functions, and molecular dynamics},}\ }\href@noop {} {\bibfield
  {journal} {\bibinfo  {journal} {Physical Review A}\ }\textbf {\bibinfo
  {volume} {2}},\ \bibinfo {pages} {975} (\bibinfo {year} {1970})}\BibitemShut
  {NoStop}%
\bibitem [{\citenamefont {Berne}\ and\ \citenamefont
  {Harp}(1970)}]{berne1970calculation}%
  \BibitemOpen
  \bibfield  {author} {\bibinfo {author} {\bibfnamefont {B.~J.}\ \bibnamefont
  {Berne}}\ and\ \bibinfo {author} {\bibfnamefont {G.}~\bibnamefont {Harp}},\
  }\bibfield  {title} {\enquote {\bibinfo {title} {On the calculation of time
  correlation functions},}\ }\href@noop {} {\bibfield  {journal} {\bibinfo
  {journal} {Advances in chemical physics}\ ,\ \bibinfo {pages} {63--227}}
  (\bibinfo {year} {1970})}\BibitemShut {NoStop}%
\bibitem [{\citenamefont {Boon}\ and\ \citenamefont
  {Yip}(1991)}]{boon1991molecular}%
  \BibitemOpen
  \bibfield  {author} {\bibinfo {author} {\bibfnamefont {J.~P.}\ \bibnamefont
  {Boon}}\ and\ \bibinfo {author} {\bibfnamefont {S.}~\bibnamefont {Yip}},\
  }\href@noop {} {\emph {\bibinfo {title} {Molecular hydrodynamics}}}\
  (\bibinfo  {publisher} {Courier Corporation},\ \bibinfo {year}
  {1991})\BibitemShut {NoStop}%
\bibitem [{\citenamefont {Daldrop}\ and\ \citenamefont
  {Netz}(2019)}]{daldrop2019mass}%
  \BibitemOpen
  \bibfield  {author} {\bibinfo {author} {\bibfnamefont {J.~O.}\ \bibnamefont
  {Daldrop}}\ and\ \bibinfo {author} {\bibfnamefont {R.~R.}\ \bibnamefont
  {Netz}},\ }\bibfield  {title} {\enquote {\bibinfo {title} {Mass-dependent
  solvent friction of a hydrophobic molecule},}\ }\href@noop {} {\bibfield
  {journal} {\bibinfo  {journal} {The Journal of Physical Chemistry B}\
  }\textbf {\bibinfo {volume} {123}},\ \bibinfo {pages} {8123--8130} (\bibinfo
  {year} {2019})}\BibitemShut {NoStop}%
\bibitem [{\citenamefont {Gao}\ and\ \citenamefont
  {Remsing}(2022)}]{Gao2022SCFNN}%
  \BibitemOpen
  \bibfield  {author} {\bibinfo {author} {\bibfnamefont {A.}~\bibnamefont
  {Gao}}\ and\ \bibinfo {author} {\bibfnamefont {R.~C.}\ \bibnamefont
  {Remsing}},\ }\bibfield  {title} {\enquote {\bibinfo {title} {Self-consistent
  determination of long-range electrostatics in neural network potentials},}\
  }\href@noop {} {\bibfield  {journal} {\bibinfo  {journal} {Nat. Commun.}\
  }\textbf {\bibinfo {volume} {13}},\ \bibinfo {pages} {1572} (\bibinfo {year}
  {2022})}\BibitemShut {NoStop}%
\bibitem [{\citenamefont {Dhattarwal}, \citenamefont {Gao},\ and\ \citenamefont
  {Remsing}(2023)}]{Dhattarwal2023SCFNN}%
  \BibitemOpen
  \bibfield  {author} {\bibinfo {author} {\bibfnamefont {H.~S.}\ \bibnamefont
  {Dhattarwal}}, \bibinfo {author} {\bibfnamefont {A.}~\bibnamefont {Gao}}, \
  and\ \bibinfo {author} {\bibfnamefont {R.~C.}\ \bibnamefont {Remsing}},\
  }\bibfield  {title} {\enquote {\bibinfo {title} {Dielectric saturation in
  water from a long-range machine learning model},}\ }\href@noop {} {\bibfield
  {journal} {\bibinfo  {journal} {J. Phys. Chem. B}\ }\textbf {\bibinfo
  {volume} {127}},\ \bibinfo {pages} {3663--3671} (\bibinfo {year}
  {2023})}\BibitemShut {NoStop}%
\bibitem [{\citenamefont {Dhattarwal}\ and\ \citenamefont
  {Kashyap}(2024)}]{Dhattarwal2024hybrid}%
  \BibitemOpen
  \bibfield  {author} {\bibinfo {author} {\bibfnamefont {H.~S.}\ \bibnamefont
  {Dhattarwal}}\ and\ \bibinfo {author} {\bibfnamefont {H.~K.}\ \bibnamefont
  {Kashyap}},\ }\bibfield  {title} {\enquote {\bibinfo {title} {Unveiling the
  influence of ionic liquid on the interfacial structure and capacitive
  performance of water-in-salt electrolytes at graphite electrodes},}\
  }\href@noop {} {\bibfield  {journal} {\bibinfo  {journal} {J. Phys. Chem. C}\
  }\textbf {\bibinfo {volume} {128}},\ \bibinfo {pages} {536--542} (\bibinfo
  {year} {2024})}\BibitemShut {NoStop}%
\bibitem [{\citenamefont {Fedorov}\ and\ \citenamefont
  {Kornyshev}(2014)}]{fedorov2014ionic}%
  \BibitemOpen
  \bibfield  {author} {\bibinfo {author} {\bibfnamefont {M.~V.}\ \bibnamefont
  {Fedorov}}\ and\ \bibinfo {author} {\bibfnamefont {A.~A.}\ \bibnamefont
  {Kornyshev}},\ }\bibfield  {title} {\enquote {\bibinfo {title} {Ionic liquids
  at electrified interfaces},}\ }\href@noop {} {\bibfield  {journal} {\bibinfo
  {journal} {Chemical reviews}\ }\textbf {\bibinfo {volume} {114}},\ \bibinfo
  {pages} {2978--3036} (\bibinfo {year} {2014})}\BibitemShut {NoStop}%
\bibitem [{\citenamefont {Zhang}\ \emph {et~al.}(2024)\citenamefont {Zhang},
  \citenamefont {Calegari~Andrade}, \citenamefont {Goldsmith}, \citenamefont
  {Raman}, \citenamefont {Li}, \citenamefont {Piaggi}, \citenamefont {Wu},
  \citenamefont {Car},\ and\ \citenamefont {Selloni}}]{zhang2024molecular}%
  \BibitemOpen
  \bibfield  {author} {\bibinfo {author} {\bibfnamefont {C.}~\bibnamefont
  {Zhang}}, \bibinfo {author} {\bibfnamefont {M.~F.}\ \bibnamefont
  {Calegari~Andrade}}, \bibinfo {author} {\bibfnamefont {Z.~K.}\ \bibnamefont
  {Goldsmith}}, \bibinfo {author} {\bibfnamefont {A.~S.}\ \bibnamefont
  {Raman}}, \bibinfo {author} {\bibfnamefont {Y.}~\bibnamefont {Li}}, \bibinfo
  {author} {\bibfnamefont {P.~M.}\ \bibnamefont {Piaggi}}, \bibinfo {author}
  {\bibfnamefont {X.}~\bibnamefont {Wu}}, \bibinfo {author} {\bibfnamefont
  {R.}~\bibnamefont {Car}}, \ and\ \bibinfo {author} {\bibfnamefont
  {A.}~\bibnamefont {Selloni}},\ }\bibfield  {title} {\enquote {\bibinfo
  {title} {Molecular-scale insights into the electrical double layer at
  oxide-electrolyte interfaces},}\ }\href@noop {} {\bibfield  {journal}
  {\bibinfo  {journal} {Nature communications}\ }\textbf {\bibinfo {volume}
  {15}},\ \bibinfo {pages} {10270} (\bibinfo {year} {2024})}\BibitemShut
  {NoStop}%
\bibitem [{\citenamefont {Zhang}\ \emph {et~al.}(2025)\citenamefont {Zhang},
  \citenamefont {Yu}, \citenamefont {Car},\ and\ \citenamefont
  {Selloni}}]{zhang2025tuning}%
  \BibitemOpen
  \bibfield  {author} {\bibinfo {author} {\bibfnamefont {C.}~\bibnamefont
  {Zhang}}, \bibinfo {author} {\bibfnamefont {Z.}~\bibnamefont {Yu}}, \bibinfo
  {author} {\bibfnamefont {R.}~\bibnamefont {Car}}, \ and\ \bibinfo {author}
  {\bibfnamefont {A.}~\bibnamefont {Selloni}},\ }\bibfield  {title} {\enquote
  {\bibinfo {title} {Tuning water dissociation at oxide--electrolyte interfaces
  with electric fields},}\ }\href@noop {} {\bibfield  {journal} {\bibinfo
  {journal} {Proceedings of the National Academy of Sciences}\ }\textbf
  {\bibinfo {volume} {122}},\ \bibinfo {pages} {e2505929122} (\bibinfo {year}
  {2025})}\BibitemShut {NoStop}%
\bibitem [{\citenamefont {Bui}\ and\ \citenamefont
  {Cox}(2025)}]{bui2025learning}%
  \BibitemOpen
  \bibfield  {author} {\bibinfo {author} {\bibfnamefont {A.~T.}\ \bibnamefont
  {Bui}}\ and\ \bibinfo {author} {\bibfnamefont {S.~J.}\ \bibnamefont {Cox}},\
  }\bibfield  {title} {\enquote {\bibinfo {title} {Learning classical density
  functionals for ionic fluids},}\ }\href@noop {} {\bibfield  {journal}
  {\bibinfo  {journal} {Physical Review Letters}\ }\textbf {\bibinfo {volume}
  {134}},\ \bibinfo {pages} {148001} (\bibinfo {year} {2025})}\BibitemShut
  {NoStop}%
\bibitem [{\citenamefont {Aguado}\ \emph {et~al.}(2003)\citenamefont {Aguado},
  \citenamefont {Bernasconi}, \citenamefont {Jahn},\ and\ \citenamefont
  {Madden}}]{aguado2003multipoles}%
  \BibitemOpen
  \bibfield  {author} {\bibinfo {author} {\bibfnamefont {A.}~\bibnamefont
  {Aguado}}, \bibinfo {author} {\bibfnamefont {L.}~\bibnamefont {Bernasconi}},
  \bibinfo {author} {\bibfnamefont {S.}~\bibnamefont {Jahn}}, \ and\ \bibinfo
  {author} {\bibfnamefont {P.~A.}\ \bibnamefont {Madden}},\ }\bibfield  {title}
  {\enquote {\bibinfo {title} {Multipoles and interaction potentials in ionic
  materials from planewave-dft calculations},}\ }\href@noop {} {\bibfield
  {journal} {\bibinfo  {journal} {Faraday Discussions}\ }\textbf {\bibinfo
  {volume} {124}},\ \bibinfo {pages} {171--184} (\bibinfo {year}
  {2003})}\BibitemShut {NoStop}%
\bibitem [{\citenamefont {Roy}\ \emph {et~al.}(2021{\natexlab{a}})\citenamefont
  {Roy}, \citenamefont {Brehm}, \citenamefont {Sharma}, \citenamefont {Wu},
  \citenamefont {Maltsev}, \citenamefont {Halstenberg}, \citenamefont
  {Gallington}, \citenamefont {Mahurin}, \citenamefont {Dai}, \citenamefont
  {Ivanov} \emph {et~al.}}]{roy2021unraveling}%
  \BibitemOpen
  \bibfield  {author} {\bibinfo {author} {\bibfnamefont {S.}~\bibnamefont
  {Roy}}, \bibinfo {author} {\bibfnamefont {M.}~\bibnamefont {Brehm}}, \bibinfo
  {author} {\bibfnamefont {S.}~\bibnamefont {Sharma}}, \bibinfo {author}
  {\bibfnamefont {F.}~\bibnamefont {Wu}}, \bibinfo {author} {\bibfnamefont
  {D.~S.}\ \bibnamefont {Maltsev}}, \bibinfo {author} {\bibfnamefont
  {P.}~\bibnamefont {Halstenberg}}, \bibinfo {author} {\bibfnamefont {L.~C.}\
  \bibnamefont {Gallington}}, \bibinfo {author} {\bibfnamefont {S.~M.}\
  \bibnamefont {Mahurin}}, \bibinfo {author} {\bibfnamefont {S.}~\bibnamefont
  {Dai}}, \bibinfo {author} {\bibfnamefont {A.~S.}\ \bibnamefont {Ivanov}},
  \emph {et~al.},\ }\bibfield  {title} {\enquote {\bibinfo {title} {Unraveling
  local structure of molten salts via x-ray scattering, raman spectroscopy, and
  ab initio molecular dynamics},}\ }\href@noop {} {\bibfield  {journal}
  {\bibinfo  {journal} {The Journal of Physical Chemistry B}\ }\textbf
  {\bibinfo {volume} {125}},\ \bibinfo {pages} {5971--5982} (\bibinfo {year}
  {2021}{\natexlab{a}})}\BibitemShut {NoStop}%
\bibitem [{\citenamefont {Roy}\ \emph {et~al.}(2020)\citenamefont {Roy},
  \citenamefont {Wu}, \citenamefont {Wang}, \citenamefont {Ivanov},
  \citenamefont {Sharma}, \citenamefont {Halstenberg}, \citenamefont {Gill},
  \citenamefont {Abeykoon}, \citenamefont {Kwon}, \citenamefont {Topsakal}
  \emph {et~al.}}]{roy2020structure}%
  \BibitemOpen
  \bibfield  {author} {\bibinfo {author} {\bibfnamefont {S.}~\bibnamefont
  {Roy}}, \bibinfo {author} {\bibfnamefont {F.}~\bibnamefont {Wu}}, \bibinfo
  {author} {\bibfnamefont {H.}~\bibnamefont {Wang}}, \bibinfo {author}
  {\bibfnamefont {A.~S.}\ \bibnamefont {Ivanov}}, \bibinfo {author}
  {\bibfnamefont {S.}~\bibnamefont {Sharma}}, \bibinfo {author} {\bibfnamefont
  {P.}~\bibnamefont {Halstenberg}}, \bibinfo {author} {\bibfnamefont {S.~K.}\
  \bibnamefont {Gill}}, \bibinfo {author} {\bibfnamefont {A.~M.}\ \bibnamefont
  {Abeykoon}}, \bibinfo {author} {\bibfnamefont {G.}~\bibnamefont {Kwon}},
  \bibinfo {author} {\bibfnamefont {M.}~\bibnamefont {Topsakal}},  \emph
  {et~al.},\ }\bibfield  {title} {\enquote {\bibinfo {title} {Structure and
  dynamics of the molten alkali-chloride salts from an x-ray, simulation, and
  rate theory perspective},}\ }\href@noop {} {\bibfield  {journal} {\bibinfo
  {journal} {Physical Chemistry Chemical Physics}\ }\textbf {\bibinfo {volume}
  {22}},\ \bibinfo {pages} {22900--22917} (\bibinfo {year} {2020})}\BibitemShut
  {NoStop}%
\bibitem [{\citenamefont {Roy}\ \emph {et~al.}(2021{\natexlab{b}})\citenamefont
  {Roy}, \citenamefont {Liu}, \citenamefont {Topsakal}, \citenamefont {Dias},
  \citenamefont {Gakhar}, \citenamefont {Phillips}, \citenamefont {Wishart},
  \citenamefont {Leshchev}, \citenamefont {Halstenberg}, \citenamefont {Dai}
  \emph {et~al.}}]{roy2021holistic}%
  \BibitemOpen
  \bibfield  {author} {\bibinfo {author} {\bibfnamefont {S.}~\bibnamefont
  {Roy}}, \bibinfo {author} {\bibfnamefont {Y.}~\bibnamefont {Liu}}, \bibinfo
  {author} {\bibfnamefont {M.}~\bibnamefont {Topsakal}}, \bibinfo {author}
  {\bibfnamefont {E.}~\bibnamefont {Dias}}, \bibinfo {author} {\bibfnamefont
  {R.}~\bibnamefont {Gakhar}}, \bibinfo {author} {\bibfnamefont {W.~C.}\
  \bibnamefont {Phillips}}, \bibinfo {author} {\bibfnamefont {J.~F.}\
  \bibnamefont {Wishart}}, \bibinfo {author} {\bibfnamefont {D.}~\bibnamefont
  {Leshchev}}, \bibinfo {author} {\bibfnamefont {P.}~\bibnamefont
  {Halstenberg}}, \bibinfo {author} {\bibfnamefont {S.}~\bibnamefont {Dai}},
  \emph {et~al.},\ }\bibfield  {title} {\enquote {\bibinfo {title} {A holistic
  approach for elucidating local structure, dynamics, and speciation in molten
  salts with high structural disorder},}\ }\href@noop {} {\bibfield  {journal}
  {\bibinfo  {journal} {Journal of the American Chemical Society}\ }\textbf
  {\bibinfo {volume} {143}},\ \bibinfo {pages} {15298--15308} (\bibinfo {year}
  {2021}{\natexlab{b}})}\BibitemShut {NoStop}%
\bibitem [{\citenamefont {Maltsev}\ \emph {et~al.}(2024)\citenamefont
  {Maltsev}, \citenamefont {Driscoll}, \citenamefont {Zhang}, \citenamefont
  {Neuefeind}, \citenamefont {Reinhart}, \citenamefont {Agca}, \citenamefont
  {Ray}, \citenamefont {Halstenberg}, \citenamefont {Aziziha}, \citenamefont
  {Schorne-Pinto} \emph {et~al.}}]{maltsev2024transient}%
  \BibitemOpen
  \bibfield  {author} {\bibinfo {author} {\bibfnamefont {D.~S.}\ \bibnamefont
  {Maltsev}}, \bibinfo {author} {\bibfnamefont {D.~M.}\ \bibnamefont
  {Driscoll}}, \bibinfo {author} {\bibfnamefont {Y.}~\bibnamefont {Zhang}},
  \bibinfo {author} {\bibfnamefont {J.~C.}\ \bibnamefont {Neuefeind}}, \bibinfo
  {author} {\bibfnamefont {B.}~\bibnamefont {Reinhart}}, \bibinfo {author}
  {\bibfnamefont {C.}~\bibnamefont {Agca}}, \bibinfo {author} {\bibfnamefont
  {D.}~\bibnamefont {Ray}}, \bibinfo {author} {\bibfnamefont {P.~W.}\
  \bibnamefont {Halstenberg}}, \bibinfo {author} {\bibfnamefont
  {M.}~\bibnamefont {Aziziha}}, \bibinfo {author} {\bibfnamefont
  {J.}~\bibnamefont {Schorne-Pinto}},  \emph {et~al.},\ }\bibfield  {title}
  {\enquote {\bibinfo {title} {Transient covalency in molten uranium (iii)
  chloride},}\ }\href@noop {} {\bibfield  {journal} {\bibinfo  {journal}
  {Journal of the American Chemical Society}\ }\textbf {\bibinfo {volume}
  {146}},\ \bibinfo {pages} {21220--21224} (\bibinfo {year}
  {2024})}\BibitemShut {NoStop}%
\end{thebibliography}%

\end{document}